\documentclass{sig-alternate-10pt}
\usepackage{times}
\usepackage{multirow} 
\usepackage{booktabs}
\usepackage{amsmath}
\usepackage{multirow} 
\usepackage{rotating}
\usepackage{tabularx}
\usepackage{booktabs}
\usepackage{graphicx}
\usepackage{subfigure}
\usepackage{pgf}
\usepackage{tikz}
\usepackage{pifont}
\usepackage{ctable}
\usepackage{alltt}
\usepackage{url}
\usepackage{wasysym}
\usepackage{wasysym}
\usepackage{comment}
\usepackage{color}
\usepackage{xcolor}
\usepackage{framed}
\usepackage{comment}
\definecolor{mydarkblue}{rgb}{0,0.8,0.45}
\definecolor{shadecolor}{rgb}{0.1,0.6,0.2}
\specialcomment{answer}{\begin{shaded}}{\end{shaded}}
\newcommand{\parag}[1]{{\noindent\bf#1.}~}

{\bfseries}{\itshape}
{\bfseries}{\itshape}
{\bfseries}{\itshape}
{\bfseries}{\itshape}
{\bfseries}{\itshape}

\begin{document}

\title{Decryptable to Your Eyes: Visualization of Security Protocols at the User Interface}
\numberofauthors{3}
\author{The Authors}
\author{
\alignauthor
DaeHun Nyang \\
\affaddr{Inha University}\\
\affaddr{Incheon, Korea}
\and 
Abedelaziz Mohaisen\thanks{Corresponding author. E-mail: mohaisen@cs.umn.edu}\\
\affaddr{University of Minnesota}\\
\affaddr{Minneapolis, Minnesota 55455}
\and 
Taekyoung Kwon\\
\affaddr{Sejong University}\\
\affaddr{Seoul, Korea}
\and 
Brent Kang \\
\affaddr{George Mason University}\\
\affaddr{Fairfax, Virginia 22030}
\and
Angelos Stavrou\\
\affaddr{George Mason University}\\
\affaddr{Fairfax, Virginia 22030}
}
\maketitle
\newcommand{\func}[1]{\textcolor{black}{#1}}
\newcommand{\id}{{\sf ID}}
\newcommand{\sig}{{\sf Sig}}
\newcommand{\ver}{{\sf Ver}}
\newcommand{\qre}{{\sf QREnc}}
\newcommand{\qrd}{{\sf QRDec}}
\newcommand{\dec}{{\sf Decr}}
\newcommand{\enc}{{\sf Encr}}
\newcommand{\fun}[1]{\textcolor{blue}{#1}}
\newcommand{\var}[1]{\textcolor{teal}{#1}}
\newcommand{\entit}[1]{\textcolor{black}{\textbf{\texttt{#1}}}}
\newcommand{\user}{\entit{user}}
\newcommand{\server}{\entit{server}}
\newcommand{\terminal}{\entit{terminal}}
\newcommand{\smartphone}{\entit{smartphone}}
\newcommand{\vpi}{\var{pi}}
\newcommand{\vpkid}{\var{pkid}}
\newcommand{\vqrekbd}{\var{qrekbd}}
\newcommand{\vekbd}{\var{ekbd}}
\newcommand{\vtrue}{\var{true}}
\newcommand{\vid}{\var{id}}
\newcommand{\vpw}{\var{pw}}
\newcommand{\fview}{\fun{view}}
\newcommand{\ba}{$\bullet$}
\newcommand{\ca}{$\circ$}
\newcommand{\baa}{\ba\ba\ba\ba\ba\ba\ba\ba\ba\ba}
\newcommand{\bab}{\ba\ba\ba\ba\ba\ba\ba\ba\ba}
\newcommand{\bac}{\ba\ba\ba\ba\ba\ba\ba\ba}

\begin{abstract}
The design of authentication protocols, for online banking services 
 in particular and any service that is of sensitive nature in
  general, is quite challenging. Indeed, enforcing security guarantees
  has overhead thus imposing additional computation and design
  considerations that do not always meet usability and user
  requirements.  On the other hand, relaxing assumptions and rigorous
  security design to improve the user experience can lead to security
   breaches that can harm the users' trust in the system.

  In this paper, we demonstrate how careful visualization design can
  enhance not only the security but also the usability of the
  authentication process.  To that end, we propose a family of
  visualized authentication protocols, a visualized transaction
  verification, and a ``decryptable to your eyes only''
  protocol. Through rigorous analysis, we verify that our protocols
  are immune to many of the challenging authentication attacks
  applicable in the literature. Furthermore, using an extensive case
  study on a prototype of our protocols, we highlight the potential of
  our approach for real-world deployment: we were able to achieve a
  high level of usability while satisfying stringent security
  requirements.
\end{abstract}


\section{Introduction}\label{sec:introduction}

Threats against the electronic and digital
financial services can be classified into two major classes of
attacks: credential stealing attacks and channel breaking
attacks~\cite{HKW06}. Credentials---such as users identifiers,
passwords, and keys---can be stolen by an attacker when they are
poorly managed. For example, a poorly managed personal computer (PC)
infected with a malicious software (malware) is an easy target for
credential attackers~\cite{Yin:2007:PCS,Stone-Gross:2009}. On the
other hand, channel breaking attacks---which allow for eavesdropping
on communication between users and a financial institution---are
another form of exploitation~\cite{hopper2001secure}. While classical
channel breaking attacks can be prevented by a proper usage of
security channel such as IPSec~\cite{doraswamy2003ipsec} and SSL
(secure sockets layer)~\cite{rescorla2001ssl}, recent channel breaking
attacks are more challenging. Indeed, ``shoulder-surfing''
attacks---or those that utilize session hijacking, phishing and
pharming, visual fraudulence, and key logging--- cannot be addressed
by data encryption.

Chief among this class of attacks are
keyloggers~\cite{herley2006login,Stone-Gross:2009,Slowinska:2009:PTE}. A
keylogger is software designed to capture all user's keyboard strokes
and make use of them to impersonate a user in financial
transactions. For example, whenever a user types in her password in a
bank's sign-in box, the keylogger intercepts the password. The
threat of malware is pervasive and can be present both in personal
computers and public kiosks. There are cases where it is necessary to
perform financial transactions using a public computer. The
biggest concern is that a user's password is likely to be stolen. Even worse, keyloggers, often rootkitted, are hard to detect since they will
not show up in the task manager process list.

To mitigate the keylogger attack, a
virtual keyboard or an onscreen keyboard with random keyboard
arrangement has been introduced. Rearranging alphabets randomly  on the
keycaps can frustrate simple keyloggers. Unfortunately, the keylogger
software, which has the control over the entire PC, can easily capture
every event and read the video buffer to create a mapping between the
clicks and the new alphabet. Another approach is to use the keyboard
hooking prevention technique by perturbing the keyboard interrupt
vector table~\cite{pemmaraju2007methods}. However, this technique is
not universal and can interfere with the operating system and native
device drivers. Also, there is another avenue for event
hooking bypassing this defense rendering it ineffective.

Considering that the keylogger sees user's keystroke, this attack is
quite similar to the shoulder-surfing attack. To prevent the
shoulder-surfing attack, many graphical password schemes have been
introduced~\cite{Gao:2008:YYG,Hayashi:2008:UYI,Kumar:2007:RSU}, but
many of them are not usable in the sense that they are quite
complicated for a person to utilize them. However, the usability is as
important as the security, because users tend not to change their
online transaction experience for higher security.

It is not enough to depend only on cryptographic techniques to prevent
attacks which aim to deceive user's visual experience. Even if all
necessary information is securely delivered to a user's computer, the
attacker residing on a user's computer can easily observe and alter
the information and show a valid-looking yet deceiving
information. Human user's involvement in the security protocol is
necessary sometimes to prevent this type of attacks, but human is not
good at complicated calculations and does not have a good memory to
remember cryptographically-strong keys and signatures. Thus, the
usability is an important factor in designing a human-involving
protocol~\cite{hopper2001secure}.
 
Our direction to solving the problem is
to introduce an intermediate device that bridges a human user and a
terminal. Then, instead of the user directly invoking the regular
authentication protocol, he/she invokes a more sophisticated but
user-friendly protocol via the intermediate helping device. Every
interaction between the user and an intermediate helping device is
visualized using a Quick Response (QR) visual code. The goal is to
avoid any memorization, perform any complex calculations, or even
extensive typing. More specifically, our approach visualizes the
security process of authentication using a smartphone-aided augmented
reality. The visual involvement of users in a security protocol boosts
both the security of the protocol and is re-assuring to the user
because he feels that he plays a role in the process. To securely
implement visualized security protocols, a smartphone with a camera is
used. Instead of executing the entire security protocol on the
personal computer, part of security protocol is moved to the
smartphone. The user input is entered via the smartphone as are
further interactions, when necessary. This visualization of some part
of security protocols enhances the security greatly and offers
protection against hard-to-defend attacks such as malicious ware and 
shoulder surfing attack, while not degrading the usability.

\subsection{Scope and contributions} In this paper, we demonstrate
how visualization can enhance not only the security but also the
usability. We do so by proposing a family of visualized authentication
protocols, visualized transaction verification, and a ``decryptable to
your eyes only'' protocol. Through rigorous analysis, we show that our
protocols are immune to many of the challenging attacks applicable to
other protocols in the literature. Furthermore, using an extensive
case study on a prototype of our protocols, we highlight the potential
of our protocols in real-world deployment meeting users shortcomings
and limitations. The original contributions of this paper are as
follows:
\begin{itemize}
\item Three protocols of authentication that utilize visualization by
  means of augmented reality to provide both high security and high
  usability. We show that these protocols are secure under several
  real-world attacks, including keyloggers, malwares, and
  shoulder-surfer; three attacks that are known to be challenging on
  authentication protocols.
\item A novel protocol for transaction verification paired with a
  protocol for secure transaction processing. Both protocols offer
  advantages due to visualization both in terms of security and usability.
\item A prototype implementation in the form of an Android
  application which demonstrates the usability of our protocols in
  real-world deployment settings.
\end{itemize}
We note that our protocols are generic and can be applied to many contexts of authentication. For example, a plausible scenario of deployment could be when considering the terminal in our system as an ATM (Automated Teller Machine), public PC, among others. Furthermore, our design does not require explicit channel between the bank and the smartphone, which is desirable in some contexts; the smartphone can be replaced by any device with the needed functionality (see section~\ref{sec:model} for more details).

\subsection{Organization} The rest of this paper is organized as follows. In section~\ref{sec:model} we review the system, trust, and attacker model used in this paper. In section~\ref{sec:arauth}, we review three novel authentication protocols. In section~\ref{sec:discussion} we extended the presentation of these protocols by discussing several implementation and design issues. In section~\ref{sec:secanalysis} we analyze the security of our protocols under several potential attacks. In section~\ref{sec:exp} we report several experiments and user studies to support the usability of our protocols. In section~\ref{sec:related} we review related work from the literature. In section~\ref{sec:con} we draw concluding remarks and point out several future work directions. 

\section{System and Threat Model}\label{sec:model}
In this section we describe the system and threat models suitable to understand the work in the rest of this paper. 

\subsection{System Model}
Our system model consists of four different entities (or participants), which are a user, a smartphone, a user's terminal, and a server. The user is an ordinary human, limited by all human's limitations and shortcomings, including limited capabilities of performing complex computations or remembering sophisticated cryptographic credentials. With a user's terminal such as a desktop computer or a laptop, the user can log in a server of a financial institute (bank) for financial transactions. After a successful login by proving possession of valid credentials, the user can do financial transactions such as money transfer and bill payment.  Also, the user has a smartphone, the third system entity, which is equipped with a camera, stores a public key certificate of the server for digital signature verification. Also, the smartphone stores a public/private key pair of the user, so if there is a channel of communication between the server and the smartphone, a secure channel can be established. Finally, the server is the last system entity, which belongs to the financial institute and performs back-end operations by interacting with the user (terminal or smartphone) on behalf of the bank. 

Assuming a smartphone entity in our system is not a far-fetched assumption, since most nowadays cell phones qualify (in terms of processing and imaging capabilities) to the device used in our work. In our system, we assume that there is no direct channel between the server and the smartphone. Also, we note that in most of the protocols proposed in this paper, a smartphone does not use the communication channel---unless otherwise is explicitly stated---so a smartphone can be replaced by any device with a camera and some proper processing power such as a camera and a portable music player with camera (iPod touch, or mobile gadget with the aforementioned capabilities). 

\subsection{Trust and Attacker Models}\label{sec:attackmodel}
For the trusted entities in our system, we assume the following. First, we assume that the channel between the server and the user's terminal (or simply PC) is {\em secured} with an SSL connection, which is in fact a very realistic assumption in most settings of electronic banking systems. Second, we assume that the server is secured by any means and is immune to every attack by the attacker; hence the attacker's concern is not breaking the server but attacking the user. Finally, with respect to the keylogger attack, we assume that the keylogger always resides on the terminal. As for the attacker model, we assume a malicious attacker with high incentives of breaking the security of the system. We assume that the attacker can do any or both of the following. 
\begin{itemize}
\item The attacker has a full control over the terminal. Thus,
\begin{itemize}
\item While residing in a user's terminal, the attacker can capture user's credentials such as a password, 
a private key, and OTP (one time pad) token string.
\item The attacker can deceive a user by showing a genuinely-looking page that actually transfers money 
to the attacker's account with the captured credentials that she obtained from the compromised terminal.
\item Or, just after a user successfully gets authenticated with a valid credential, 
the attacker can hijack the authenticated session.  
\end{itemize}
\item The attacker is capable of creating a fake server which she can use to launch phishing or pharming attacks.
\end{itemize}

For the smartphone, we assume that it is always trusted and immune to compromise. It is however noted that relaxing this assumption would provide certain security guarantees depending on the protocol. For example, the assumption in protocol 1 that uses two factors can be relaxed so as not only the terminal but also smartphone is compromised (one of them at a time but not both together). However, for protocols 2 and 3, we restrict this scenario so as the smartphone is always trusted and no malware nor keyloggers can be installed on it. Notice that this assumption is in line with other assumptions made on the smartphone's trustworthiness when used in similar protocols to those presented in this paper~\cite{DBLP:journals/ijsn/McCunePR09,DBLP:conf/sp/McCunePR05,DBLP:conf/fc/ParnoKP06}. 

In our protocols, we also assume several cryptographic primitives. For example, in all protocols, we assume that users have pairs of public/private keys used for message signing and verification. In protocol 1 we assume users have passwords used for their authentication. In protocol 2 we assume that the server has the capability of generating one time pads, used for authentication. Notice that these assumptions are not far-fetched, since most banking services use such cryptographic credentials. For example, with most banking services, the use of digital certificates issued by the bank is very common. Furthermore, the use of such cryptographic credentials and maintaining them on a smartphone does not require any technical background at the user side, and is suited for wide variety of users. Further details on these credentials and their use are explained along with the specific protocol in this paper.  

\subsection{Linear and Matrix Barcodes}
\begin{figure}[htbp]
\begin{center}
\subfigure[Barcode (code 128)]{
   \includegraphics[height =1.3cm] {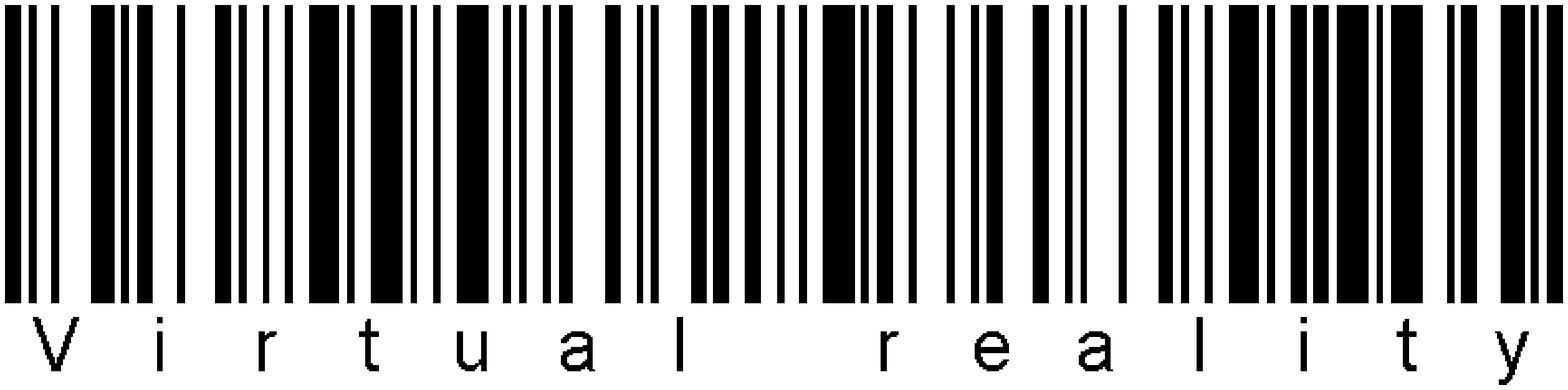}
   \label{fig:bar}
 }\\
\subfigure[QR barcode]{
   \includegraphics[height =2.5cm] {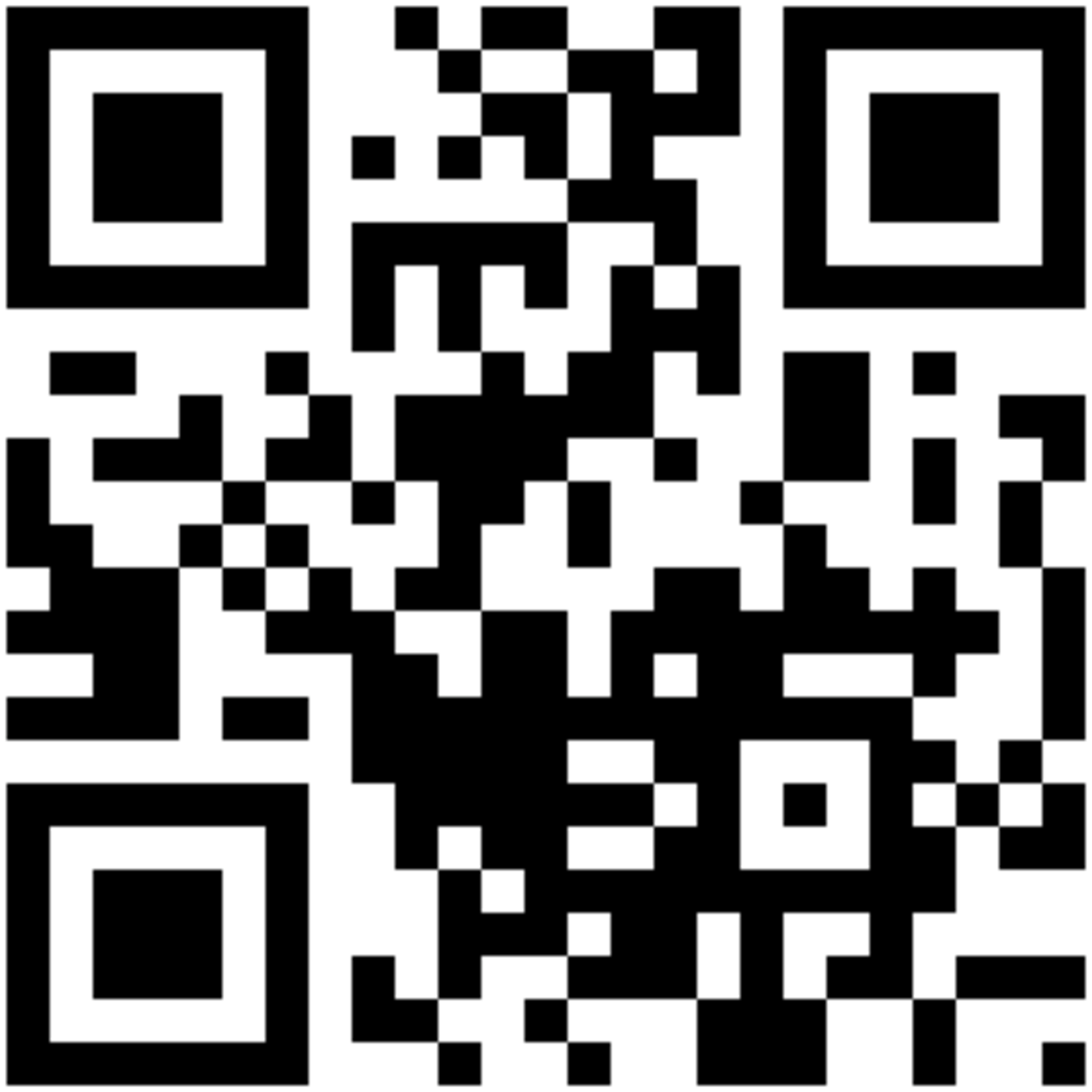}
   \label{fig:qr}
 }
\subfigure[QR barcode]{
   \includegraphics[height =2.5cm] {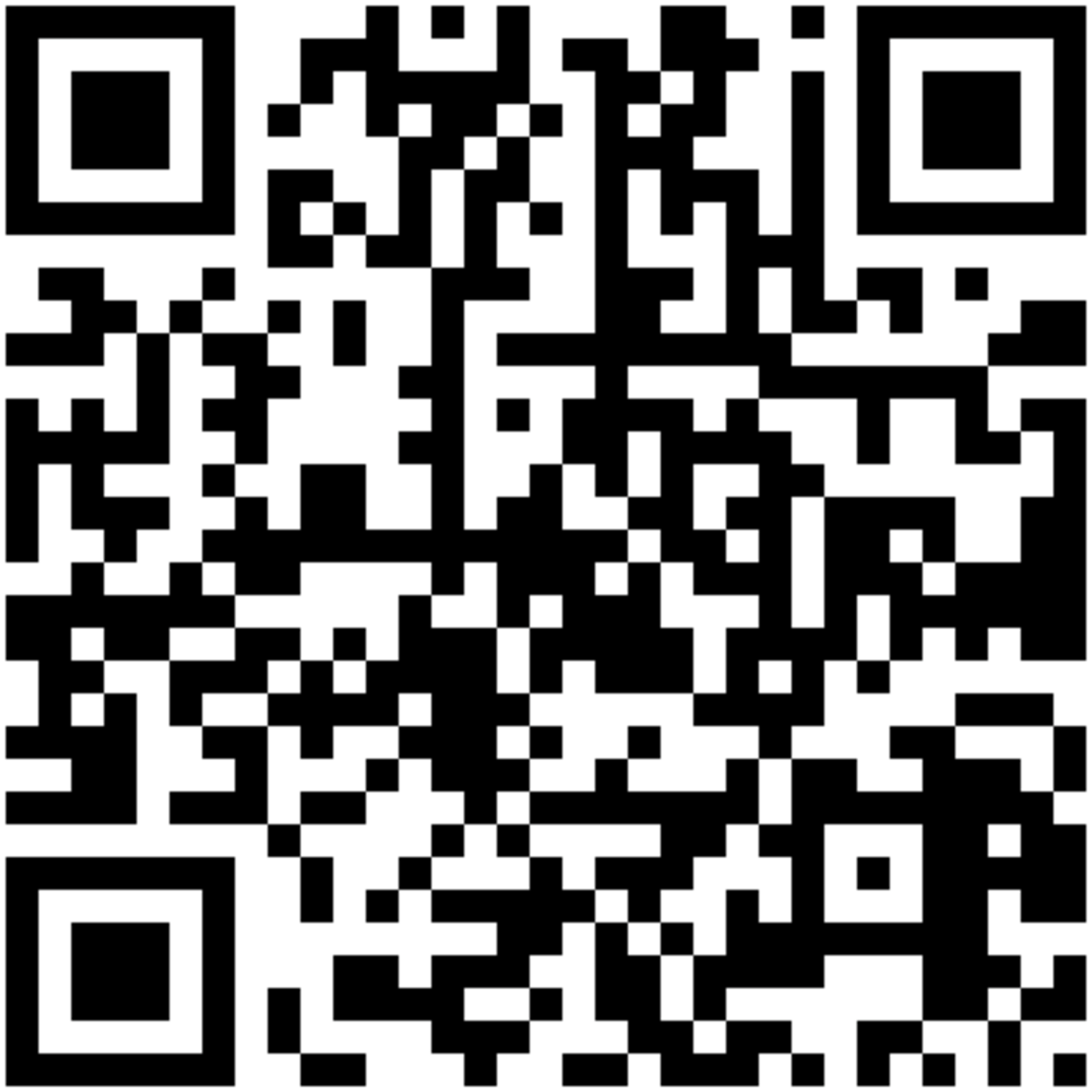}
   \label{fig:qrenc}
 }
\caption{Three different barcodes encoding the statement ``Virtual reality''. \subref{fig:bar}~is a linear barcode (code 128), and~\subref{fig:qr} and~\subref{fig:qrenc} are matrix barcodes (of the QR barcode standard). While~\subref{fig:qr} encodes the plain text, \subref{fig:qrenc} encodes an encrypted version using the AES-256 encryption algorithm in the cipher-block chaining (CBC) mode (note this last code requires a password for decryption).}
\label{fig:codes}
\end{center}
\end{figure}

A barcode is an optical machine-readable representation of data, and it is widely used in our daily life since it is attached in every type of products for identification. In the nutshell, barcodes are mainly two types: linear barcodes and matrix (or two dimensional, also known as 2D) barcodes. While linear barcodes---shown in Figure~\ref{fig:bar}---have a limited capacity, which depends on the coding technique used and can range from 10 to 22 characters, 2D barcodes---shown in Figure~\ref{fig:qr} and Figure~\ref{fig:qrenc}---have higher capacity, which can be more than 7000 characters. For example, the QR (Quick Response) barcode~\cite{qrcode}---a widely used 2D barcode---can hold 7,089 numeric, 4,296 alphanumeric, and 2,953 binary characters~\cite{qrcode}, making it a very good high-capacity candidate for storing plain and encrypted contents alike.

Both linear and matrix barcodes are popular and have been widely used in a lot of industries including, but not limited to, automotive industries, manufacturing of electronic components, and bottling industries, among many others. Thanks to their greater capacity, matrix barcodes are even proactively used for advertisement so that a user who has a smartphone can easily scan them to get some detailed information about advertised products. This model of advertisement---and other venues of using these barcodes in areas that are in touch with users---created the need for barcode's scanners developed specifically for smartphones. Accordingly, this led to the creation of many popular commercial and free barcode scanners that are available for smartphones such as iPhone and Andriod phones alike. This includes RedLaser~\cite{redlaser}, BarcodeScanner~\cite{bcscanner}, ShopSavvy~\cite{savvy}, QR App~\cite{qrapp}, among others.

\section{Secure Transactions with Augmented Reality}\label{sec:arauth}
In this section we describe thee different protocols for user authentication with augmented reality. Before getting into the details of these protocols, we review the notations of algorithms used in our system and these protocols as building blocks. Our system utilizes the following algorithms:
\begin{itemize} 
	\item ${\sf Encr}_k(\cdot)$: an encryption algorithm which takes a key $k$ and a message $M$ from set ${\mathcal{M}}$ and outputs a ciphertext $C$ in the set $\mathcal{C}$.
	\item ${\sf Decr}_k(\cdot)$: a decryption algorithm which takes a ciphertext $C$ in $\mathcal{C}$ and a key $k$, and outputs a plaintext (or message) $M$ in the set $\mathcal{M}$. 
	\item ${\sf Sign}(\cdot)$: a signature generation algorithm which takes a private key $SK$ and a message $M$ from the set $\mathcal{M}$, and outputs a signature $\sigma$. 
	\item ${\sf Verf}(\cdot)$: a signature verification algorithm which takes a public key $PK$ and a signed message $(M,\sigma)$, and returns {\tt valid} or {\tt invalid}. 
	\item ${\sf QREnc}(\cdot)$: a QR encoding algorithm which takes a string $S$ in $\mathcal{S}$ and outputs a QR barcode.
	\item ${\sf QRDec}(\cdot)$: a QR decoding algorithm which takes a QR barcode and returns a string $S$ in $\mathcal{S}$.
\end{itemize}
Any signature scheme with provable security guarantees can be used to serve the purpose of our system. For details on the notion of signatures security see~\cite{GoldwasserMR88}.
In particular, and for efficiency reasons, we recommend the short signature in~\cite{boneh2004short}. 

\subsection{An Authentication Protocol with Password and Randomized Onscreen Keyboard}
Our first protocol, which is referred to as Protocol 1 in the rest of this paper, makes use of the a password shared between the server and the user, and a randomized onscreen keyboard. A high-level event-driven code describing the protocol is shown in Figure~\ref{fig:protocol1}. A further detailed description is as follows. 
\begin{enumerate}
\item The user connects to the server and sends her $\id$. 
\item The server checks the received $\id$ to retrieve the user's public key ($PK_{ID}$) from the database.
The server prepares $\pi$, a random permutation of a keyboard arrangement, and encrypts it with the public key to obtain $E_{KBD}=\enc_{PK_{\id}} (\pi)$. Then, it encodes the ciphertext with QR encoder to obtain $QR_{E_{KBD}}=\qre(E_{k_{\id}} (\pi))$. The server sends the result with a blank keyboard.
\item In the user's browser, a QR barcode ($QR_{E_{KBD}}$) is displayed together with a blank keyboard. 
Because the onscreen keyboard does not have any alphabet on it, the user cannot input her password.
Now, the user executes her smartphone application which first decodes the QR barcode by applying $\qrd(QR_{E_{KBD}})$ 
to get the ciphertext ($E_{KBD}$).  The ciphertext is then decrypted by the smartphone application with the private key of the user to display ($\pi = \dec_{SK_{ID}} (E_{KBD})$) on the smartphone's screen.
\item When the user sees the blank keyboard with the QR barcode through an app on the smartphone 
that has a private key or a shared secret, alphabets and numbers appear on the blank keyboard 
and the user can click the proper keycap for the password. User clicks his/her password on the computer monitor
while seeing the keyboard layout through the smartphone.
\item Server checks whether the password is correct or not.
\end{enumerate}
\begin{figure}[htb]
\begin{center}
\begin{minipage}[b]{0.8\linewidth}
{\scriptsize
\begin{alltt}
01: \user::\func{user.send}(server, id)
02: \server::__upon_id_arrival:
03:     if(\server.\func{verify}(id) == true):
04:          pkid = \server.db.\func{find}(id)
05:          pi = \server.generate_random_kb()
06:          ekbd = server.encrypt(pkid, pi)
07:          qrekbd = server.qrencode(ekbd)
08:          server.send(user, qrekbd)
09: \terminal::__upon_qrekbd_arrival:
10:     terminal.view(qrekbd)
11:     terminal.view_blank_kb(pi)
12: \smartphone::__upon_qrekbd_view:
13:     qrekbd = smartphone.capture(qrekbd)
14:     ekbd = smartphone.qrdecode(qrekbd)
15:     pi = smartphone.decrypt(skid, ekbd)
16:     smartphone.view(pi)
17: \user::__upon_pi_view:
18:     pw = user.inputpassword(terminal)
19: \terminal::upon_pw_input:
20:     terminal.send(server, pw)
21: \server::__upon_pw_arrival:
22:     if(server.verify(id, pw) == true):
23:          server.authenticate(user)
24:     else:
25:          server.deny(user)
\end{alltt}
}
\end{minipage}
\caption{Event-driven high-level description of the authentication protocol with password and a randomized onscreen blank keyboard (protocol 1).}\label{fig:protocol1}\vspace{-4mm}
\end{center}
\end{figure}

\begin{figure}[htb]
\begin{center}
\begin{minipage}[b]{0.8\linewidth}
\scriptsize
\begin{alltt}
01: \user::user.send(server, id)
02: \server::__upon_id_arrival:
03:     if(server.verify(id) == true):
04:          pkid = server.db.find(id)
05:          otp = server.generate_otp()
06:          eotp = server.encrypt(pkid, otp)
07:          qreotp = server.qrencode(eotp)
08:          server.send(user, qreotp)
09: \terminal::__upon_qreotp_arrival:
10:     terminal.view(qreotp)
11:     terminal.view_kb()
12: \smartphone::__upon_qreotp_view:
13:     qreotp = smartphone.capture(qreotp)
14:     eotp = smartphone.qrdecode(qreotp)
15:     otp = smartphone.decrypt(skid, eotp)
16:     smartphone.view(pi)
17: \user::__upon_kb_view:
18:     otp = user.inputotp(terminal)
19: \terminal::upon_otp_input:
20:     terminal.send(server, otp)
21: \server::__upon_otp_arrival:
22:     if(server.verify(id, otp) == true):
23:          server.authenticate(user)
24:     else:
25:          server.deny(user)
\end{alltt}
\end{minipage}
\caption{Event-driven description of the authentication protocol with one-time pad (OTP) tokens (protocol 2).}\label{fig:protocol2}\vspace{-4mm}
\end{center}
\end{figure}

{\noindent\bf Message signing.}
For the generality of the purpose of this protocol and the following protocols, and to prevent the terminal from misrepresenting the contents generated by the server, one can establish the authenticity of the server and the contents generated by it by adding the following verification process. When the server sends the random permutation to the user, it signs the permutation using the server's private key and the resulting signature is encoded in a QR code. Before decrypting the contents, the user establishes the authenticity of the contents verifying the signature against the server's public key. Both steps are performed using the {\sf Sign} and {\sf Verf} algorithms. Verification is performed by the smartphone so as to avoid any man-in-the-middle attack by the terminal.

\subsection{Authentication With Random Strings} 
In this section we introduce an alternative protocol to that explained in the previous section. The following protocol (referred to as Protocol 2 in the the rest of the paper) relies on a strong assumption; it makes use of a random string for authentication.  A high-level event-driven code describing the protocol is shown in Figure~\ref{fig:protocol2}. A further detailed description is as follows. 
\begin{enumerate}
\item The user connects to the server and sends her $\id$. 
\item Th server checks the $\id$ to retrieve the user's public key ($PK_{ID}$) from the database. The server then picks a fresh random string $OTP$ and encrypts it with the public key to obtain $E_{OTP}=\enc_{PK_{\id}} (OTP)$. 
\item In the user's browser, the QR barcode is displayed and the page prompts the user to type in the string. 
\item The user decodes the QR barcode. Because the random string is encrypted with user's public key ($PK_{\id}$), the 
user can read the OTP string only through her smartphone and type in the $OTP$ with a physical keyboard.  
\item The server checks the result and if it matches with what the server has sent earlier, the user is authenticated. Otherwise, the user is denied.
\end{enumerate}

\subsection{Authentication Using Visual Channel for Smartphone-to-Terminal Communication}\label{upload}
Our last protocol, referred to as protocol 3 in the rest of the paper, takes advantage of the visual channel between the smartphone and the terminal. In this protocol, we further assume that the terminal is provided with a camera, which is not a far-fetched assumption.  Equipped with a camera, a terminal can easily receive the user's input from the smartphone via the QR barcode and forward it to the server. At terminal, a plugin that reads and decodes the QR barcode and sends it to the server is installed in a web browser. A high-level event-driven describing the protocol is shown in Figure~\ref{fig:protocol3}. A further detailed description is as follows.

\begin{enumerate}
\item The user connects to the server and receives a login box with a QR-encoded random nonce $N$.
\item The user executes an application on the smartphone which reads the nonce $N$ and prompts the user to input her $\id$ and password ({\sf PW}). After the user inputs her $\id$ and password on the smartphone, the application will encrypt $N$, $\id$, the password, and the server's name with the server's public key ($PK_{Srv}$), where the encryption algorithm is IND-CCA2 secure. The resulting ciphertext ($\enc_{PK_{Srv}}(N, \id, {\sf PW}, Srv)$) is then encoded to a QR barcode and displayed on the smartphone screen.
\item The user puts the QR barcode on the screen at the terminal's camera. The plugin reads and decodes the QR barcode to extract the ciphertext of the encrypted credentials and nonce. The ciphertext is sent to server.

\item The server decrypts the ciphertext and checks the nonce, $\id$ and password and decided upon their validity whether the user is authenticated or not. 
\end{enumerate}

\begin{figure}[htb]
\begin{center}
\begin{minipage}[b]{0.99\linewidth}
{\scriptsize
\begin{alltt}
01: \user::user.connect(server)
02: \server::__upon_user_connection:
03:     nonce = server.generate_nonce()
04:     qrnonce = server.qrencode(nonce)
05:     server.send(user,qrnonce)
06: \terminal::__upon_qrnonce_arrival:
07:     terminal.view(qrnonce)
08: \smartphone::__upon_qrnonce_view
09:     qrnonce = smartphone.capture(qrnonce)
10:     nonce = smartphone.qrdecode(qrnonce)
11:     smartphone.view_loginbox()
12: \user::__upon_loginbox_view:
13:     (id, pw) = user.input_credentials()
14: \smartphone::__upon_credentials_input:
15:     encredentials = smartphone.encrypt(id, pw, nonce)
16:     qrencredentials = smartphone.qrencode(encredentials)
17:     smaprtphone.view_qrenc(qrencredentials)
18: \terminal::__upon_qrenc_view:
19:     qrencredentials = terminal.capture(qrencredentials)
20:     encredentials = terminal.qrdecode(qrencredentials)
21:     terminal.send(server, encredentials)
22: \server::__upon_encredentials_arrival:
23:     (id, pw, nonce) = server.decrypt(encredentials)
24:     if(server.valid(user, nonce) = true)
25:         if(server.check(id, pw) = true):
26:             authenticate(user)
27:         else:
28:             deny(user)
29:     else:
30:         deny(user)
\end{alltt}
}
\end{minipage}
\caption{Event-driven high-level description of the authentication protocol with visual channel for user-terminal communication (Protocol 3).}\label{fig:protocol3}\vspace{-4mm}
\end{center}
\end{figure}
The idea of using QR barcode on the phone's screen as an upload channel (from phone to terminal) can be used in many other schemes~\cite{Lim07}. Instead of using cellular network for a phone to send the confirmation, we can use this QR barcode-based channel from phone to a terminal.  

The use of the visual channel to input encrypted credentials from the smartphone to the terminal has an interesting security implication when considered in context. As it is the case with using e-banking on untrusted terminals (say, in public library) imagine that such terminal is infected with a virus,  or has a malware, which could be a keylogger that stores the credentials if the user is to input them directly on the terminal. If the user is to use the login credentials directly on the terminal, it is obvious that these credentials will be compromised. On the other hand, if these credentials are to be transferred using the visual channel between the smartphone and the terminal, chances for logging these credentials on the terminal by the keylogger are obsolete. 
\section{Discussion}\label{sec:discussion}
Some of the technical issues in the three protocols that we have introduced in the previous section call for further discussions and clarification. In this section, we elaborate on how to handle several issues related to our protocols, such as session hijacking, transaction verification, and securing transactions.

\subsection{Prevention of Session Hijacking}
Even with secure authentication, an attacker controlling entities in the system participating in the authentication process---the terminal in particular---via a malware can hijack the authentication session just immediately after a user inputs her correct password or an OTP. To detect the session hijacking, the smartphone---triggered by the user---may send transaction-related information to the server via a side channel such as the cellular network or the visual channel introduced in section~\ref{upload}. Just after running the password input procedure in section~\ref{sec:arauth}, the smartphone application may send additional information on the user's transaction request that is signed by the smartphone's private signing key and/or encrypted with the server's public key through cellular network. By adding this step, a server can be sure that the critical transaction request from the user's terminal is not altered.

\subsection{Transaction Verification}
Visual fraudulence, such as man-in-the-browser attack~\cite{Karlof:2007:DPA}, phishing attack~\cite{Dhamija:2005:BAP}, and the pharming attack~\cite{Karlof:2007:DPA,Crites:2008:OES}, is an attack to show valid looking HTML pages, but to perform some malicious behaviors either by malicious code in the HTML page or by  making user visit malicious sites. The visual fraudulence is quite effective and critical if conducted in financial transaction scenarios. The nature of the visual fraudulent attack is to deceive people's view and thus it is not easy to devise an effective and useable method to prevent it. Also, situations are worse in practical contexts noting that the malicious ware performing the man-in-the-browser attack can be easily implemented in a form of a browser helper object.

Assume that a given user is visiting a banking server and is about to transfer money to other account. Even if the user's terminal is infected with some malicious ware or the user is visiting a phishing site, she cannot recognize it easily, because the HTML page that the user is watching is visually the same as the genuine page. Even when a bank server asks the user to input credentials such as a password, a one time password generated from a hardware token, and a certificate based signature to confirm the transaction, the user is willing to input her credentials as requested, and with the credential information the attacker is able to prepare a valid transfer request to her account. In this section, we show an effective and useable approach to defeat these visual fraudulence with the aid of a smartphone. Similar in essence to the previous discussed protocols, our protocol for preventing the visual fraudulence consists of the following:

\begin{enumerate}
\item The user sends an HTML page to request money transfer, for instance. This page might include a receiver's account number, the amount of money to be transferred, and the receiver's name, etc.
\item The server responds with confirmation HTML page along with a QR code that includes transaction information and a digital signature of messages in the HTML page. The page might include a code to prompt a user to input credentials to confirm the order.
\item The user reviews the HTML page received from the server then---with an application on her smartphone---she takes a snapshot of the QR barcode.
\item The application on the smartphone decodes and verifies the digital signature over the attached message with the server's public key. If the signature is valid, it will show the message with a mark indicating the signature was verified. Otherwise, it will warn the user with a mark indicating invalid signature. The mark may be a background color (green on a valid signature and red on an invalid one) or simple warning words. 
\item The user checks if the message matches with the one in the HTML page and the signature is verified. If it is valid and correctly verified, the user continues to confirm the transaction by inputting her credential.
\end{enumerate}

\subsection{Securing Transactions}
Financial transactions are usually secured by encrypting all transaction-related information during the transmission. In many cases, the encrypted information should be decrypted at the terminal (user's PC, most likely, or a PC at public place) to be shown to the user. However, under the assumption that there is a malware inside the terminal, the attacker does not need to break the cipher, but is enough to read the information after being correctly decrypted. The encrypted channel is established just between the server and the user's terminal. To make transactions more secure, it is needed to extend the encrypted channel beyond the user's terminal. Accordingly, instead of decrypting the ciphertext at the user's
terminal, we will decrypt it at the smartphone. Steps of securing the transaction include the following:
\begin{enumerate}
\item The server prepares an HTML document that has encrypted data and their corresponding QR barcodes.  The ciphertext is encrypted with the user's public key.
\item At the user's terminal, the document is shown but not decrypted. The user executes an application on the smartphone and sees through it the HTML page. The application captures the QR barcodes in the HTML document and then decodes and decrypts them. On the smartphone's screen, the HTML document with decrypted information will be viewed.
\end{enumerate}
The sensitive data might be transaction-related account numbers, balance, account holder's name, and ID, among others.

\section{Security Analysis}\label{sec:secanalysis}
In this section we analyze the security of our scheme under several attack scenarios and show how these attacks are defended. In particular, we consider brute-forcing attacks, keyloggers, malicious ware, and shoulder surfing attacks. 

\subsection{Key Space and Brute-Force Attacker}
In our protocols, several stages include encryption of sensitive information such as credentials, which are of interest to the attacker (including the user $\id$, password, and nonce generated by the server). In our prototype, and system recommendations for wide use of our protocols, as well as the description provide above for the different protocols, we consider public key cryptography. Furthermore, we suggest key length that provides good security guarantees. This includes the use of RSA-2048, which is very infeasible to attack using the most efficient brute-force attack. This applies to both encryption and signature algorithms used in the protocols.

Notice that all public key cryptography in our protocols (except for signing and verification) can be replaced by symmetric key cryptography, which is far more efficient (despite that computations in our protocols are marginal). Furthermore, such replacement of cryptographic techniques will not affect the security guarantees of our protocols if standard algorithms and key length are used---e.g. AES 192, which is infeasible to brute-force. In our prototype, we use the latter symmetric key cryptography for securing communication. 

\subsection{Keyloggers}\label{sec:keylogger} The keylogger is a small piece of malicious software that logs all keystrokes input by the user, which could potentially include authentication credentials, and forward them to the attacker. This type of malware is popular and widely reported in many contexts~\cite{Yin:2007:PCS,herley2006login,Stone-Gross:2009,Slowinska:2009:PTE,Holz:2009:LMU}. In our protocols, input is expected by the user, and in every protocol one or another type of input is required. Our protocols---while designed with the limitations and shortcoming of users in mind, and aim at easing the authentication process by means of visualization---are aimed explicitly at defending against the keylogger attacks. Here, we further elaborate on the potential of using keyloggers as an attack and the way they impact each of the three protocols. While we believe that having a keylogger installed on the smartphone is hard to achieve, we discuss when it is  hypothetically possible and show how this affects the security of our protocols. Narrowing the discussion to the case where keyloggers can be only installed on the terminal strengthens all of the arguments.

\parag{Protocol 1} In the first protocol, a randomized blank keyboard is posted on the terminal whereas another keyboard with the alphanumerics on it is posted on the smartphone. Because the protocol does not require a user any keyboard input on the smart phone side, the protocol is immune against the keylogger attack. User just checks the keyboard layout on the phone and there is not input from a user. Obviously, the terminal might be compromised, but the keylogger will be able to only capture what keystrokes are used on blank keyboard. Thus, the keylogger will not be able to know which alphanumeric character is being used.

\parag{Protocol 2} Authentication in this protocol is solely based on a random string generated by the server. The random string is encrypted by the public key of the user, and verified against her private key. The main objective of using OTP is that it is for one time use. Accordingly, if the keylogger is installed on the terminal, the attacker obviously will be able to know the OTP but will not be able to reuse it for future authentication. Alternatively, a keylogger installed on the smartphone will not be able to log any credentials, since no credentials are typed in. It is worth noting that the attacker may try to block users from being authenticated and reuse the OTP immediately. In this case, mitigations explained in section~\ref{sec:discussion} can be used to remedy the attack.

\parag{Protocol 3} In the third protocol, authentication is established based on credentials provided by the user on her smartphone. Accordingly, having a keylogger on the terminal will not enable the attacker to obtain any credentials required for authentication. On the other hand, if the keylogger is installed on the smartphone, a relatively far-fetched assumption, the keylogger will be able to log the password and user name. By themselves alone, however, both it is not sufficient for the attacker to be authenticated since the successful authentication requires additionally knowing the user's private key, which is necessary to obtain the nonce sent by the server and to compute the authentication response. 

\subsection{Malicious Software (malware)}
The term malware is generic, and is technically used to describe any type of code with malicious intentions, including keyloggers. 
It is obvious that an attacker who successfully compromised a smart phone that has a private key that is a whole credential required to break the system in protocol 2 and 3 will be always successful to break the systems except the protocol 1 that requires both password and private key. 

\subsection{Shoulder-Surfing Attacks}
The shoulder surfing is a powerful attack in the context of password-based authentication and human identification~\cite{vizcaino1994method,hopper2001secure,kumar2007reducing}. In this attack, the attacker tries to know credentials, such as passwords or PINs (personal identification numbers) by stealthily looking over the shoulder of a user inputting these credentials into the systems. These attacks are powerful and efficient in crowded places.  

In our first protocol, observing the terminal or the smartphone keyboard layout (on the smartphone screen) alone would not reveal the credentials of the user. Observing both at the same time in a shoulder surfing attack, and mapping stroked keys on the terminal to those on the smartphone screen would reveal the credentials of the user. However, being able to successfully launch this attack is a non-trivial task, and requires the attacker to be in a position very close to the user, which would raise suspicions of the user. 

One may argue that requiring user caution to the potential of the shoulder-surfing attack contradicts the usability arguments of our protocols we advocated in this paper. Especially, the opposing argument might sound appealing given that an attacker might be equipped with vision-enhancing devices, such as cameras, to improve chances of the attack. In response, we argue two issues. First, to succeed in this attack, the attacker needs to have a good resolution-camera, and still be in proximity of the user, to capture passwords being used. Second, users who are aware that the
terminal they are avoiding is potentially compromised would be highly likely cautious for susceptible actions around them---such as that in the first issue above, which does not thwart the usability arguments we claim for our protocols.

In the second protocol, OTP tokens are used for authentications. OTP tokens are one-time used, providing high entropy, and are human-unfriendly making them hard to remember and recall. Accordingly, a shoulder surfer would not benefit from launching an attack by trying to observe what the user at the terminal is inputting. Last, in the third protocol input is performed at the smartphone, and shoulder surfing on the terminal will not benefit the attacker. Same as in the first protocol, the ability of successfully launching an attack by shoulder-surfing would require the attacker to be in very good proximity from the user, which would raise the user's suspicions about the intentions of the attacker.
 
\subsection{Comparison}
To sum up, we compare the three protocols and the way they perform against several attacks. We consider the scenarios where the attacker has control over either the terminal or smartphone---see above. The comparison is in Table~\ref{tab:comp}.

\begin{table}
\begin{center}
\caption{A comparison of the three protocols and their resistance to different attacks when the terminal and the smartphone are under control of the attacker.}\label{tab:comp}
\begin{tabular}{rccccc}
\toprule
Attack & {brute-force} 	& {keyloggers} 	& {malware} 	& {surfer} \\
\midrule
\multicolumn{5}{c}{\bf Protocol 1: Onscreen randomized keyboard}  \\
\midrule
Smartphone	& \ding{52} 	&\ding{52} 	&\ding{52} 	&\ding{52} \\
Terminal		& \ding{52} 	&\ding{52} 	&\ding{52} 	&\ding{52} \\
\midrule
\multicolumn{5}{c}{\bf Protocol 2: OTP tokens}  \\
\midrule
Smartphone 	& \ding{52} 	&\ding{52} 	&\ding{54} 	&\ding{52} \\
Terminal	 	& \ding{52}	&\ding{52} 	&\ding{52} 	&\ding{52} \\
\midrule
\multicolumn{5}{c}{\bf Protocol 3: Visual channels for authentication}  \\
\midrule
Smartphone 	& \ding{52} 	&\ding{54} 	&\ding{54} 	&\ding{52} \\
Terminal	 	& \ding{52} 	&\ding{52} 	&\ding{52} 	&\ding{52} \\
\bottomrule
\end{tabular}
\end{center}
\end{table}

\section{Implementation and User Study}\label{sec:exp}
\begin{figure*}[htb]
\centering
\subfigure[{Terminal Keyboard}]{\label{fig:1}\includegraphics[height=3.2cm]{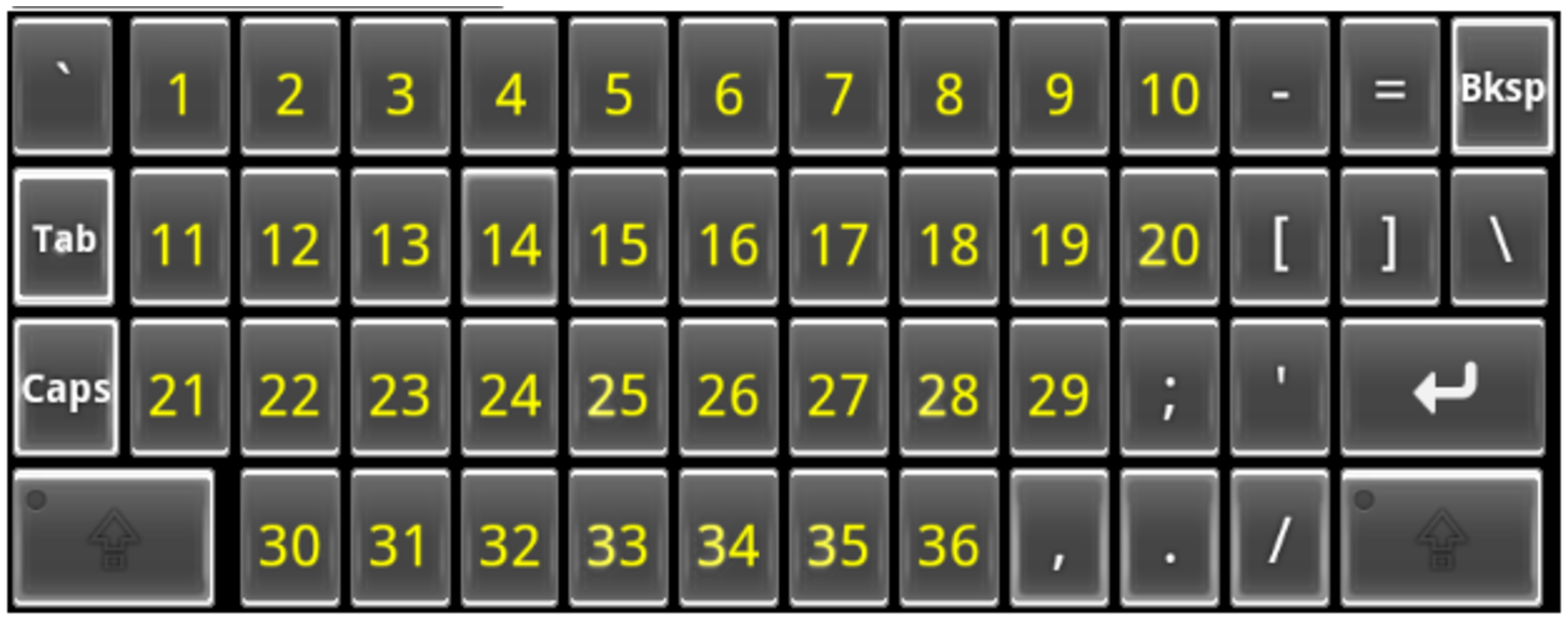}}
\subfigure[{QR code}]{\label{fig:2}\includegraphics[height=3.2cm]{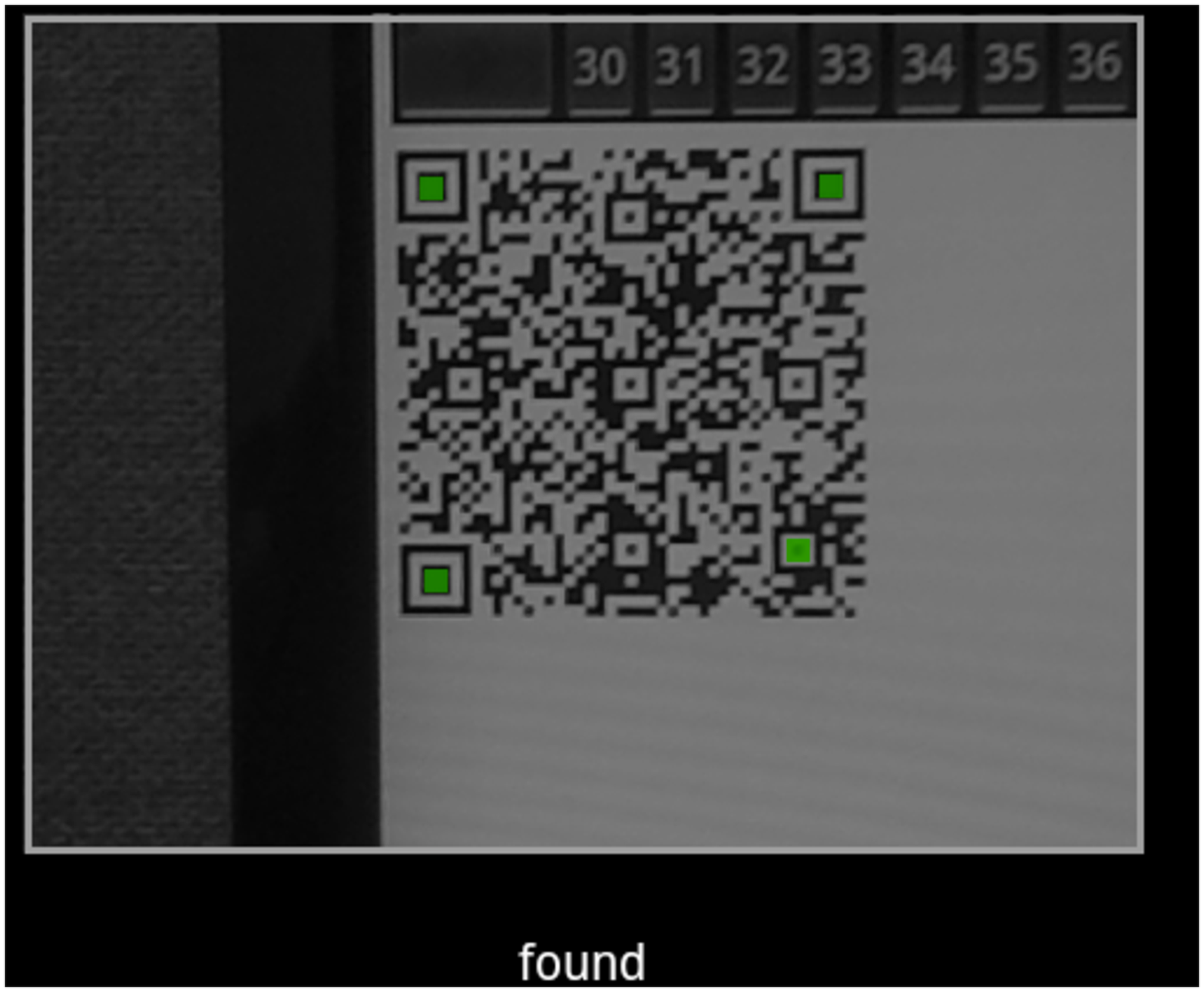}}
\subfigure[{Smartphone Keyboard}]{\label{fig:3}\includegraphics[height=3.2cm]{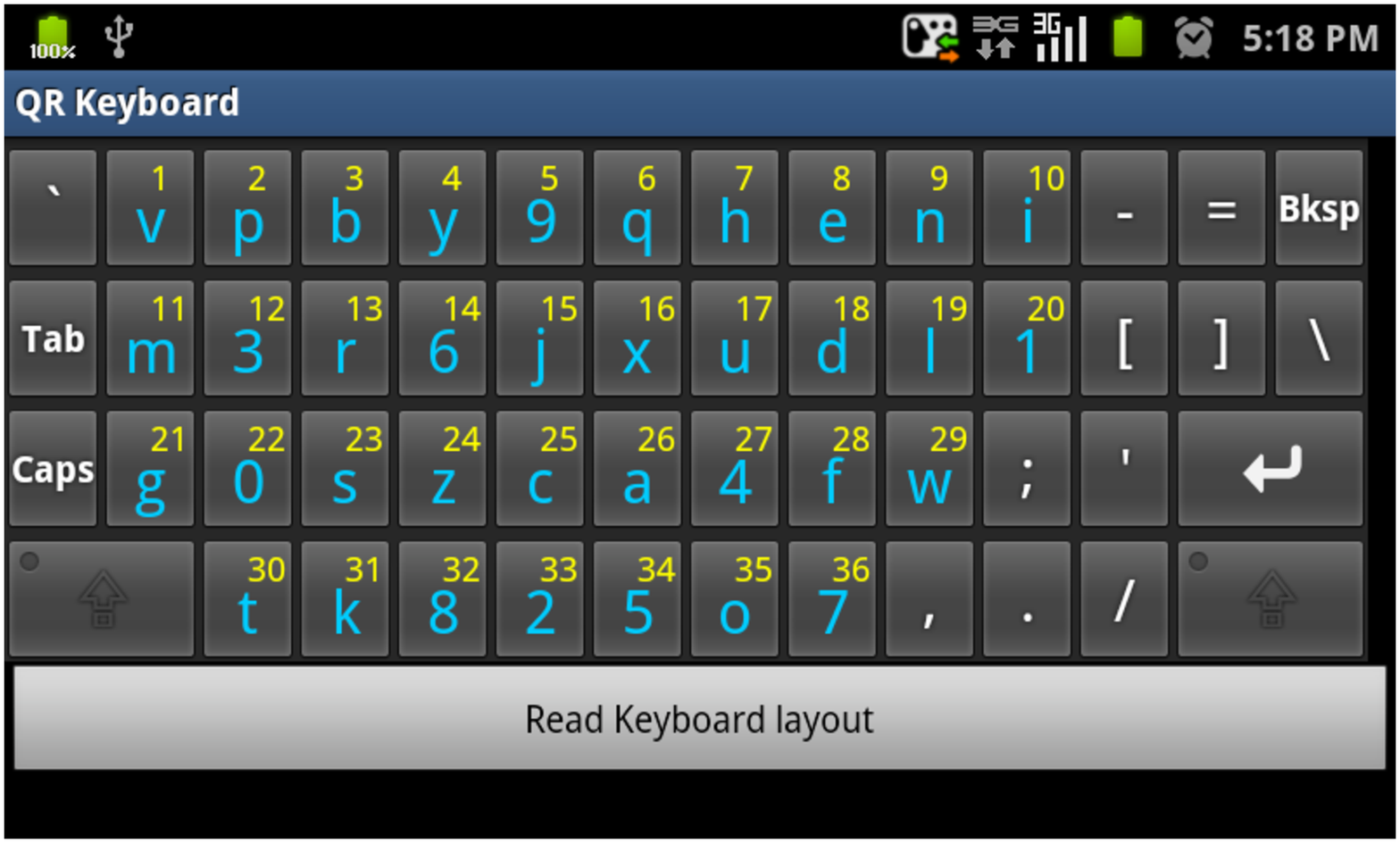}}
\caption{Snapshots of the prototype we developed to demonstrate our authentication protocols. Fig.~\subref{fig:1}. shows the blank keyboard posted at the terminal, where numerics are used to ease the process of input by the user. Fig.~\subref{fig:2} shows the QR barcode on the terminal as being captured and recognized by the smartphone application. Fig.~\subref{fig:3} shows the decoded randomized layout of the keyboard obtained from the QR barcode after decryption as viewed on smartphone.}
\label{fig:prototype}
\end{figure*}

We developed a prototype of our protocols as an Android application. The application can run on any smartphone with Android OS~\cite{android} (version 2.2 or later). Our application uses the UTF-8 standard~\cite{utf8} for the keyboard encoding, AES-192 encryption algorithm~\cite{aes} (in the counter mode) for contents encryption, Base64 encoding~\cite{base64} for byte-to-character encoding of encrypted contents, and uses the Google API for creating and reading the QR barcodes. For that, we particularly use ZXing~\cite{zxing}, an open source implementation for reading several standards of the 1D and 2D barcodes. Snapshots of the Android application, the terminal, and QR barcode in the prototype that we developed are shown in Figure~\ref{fig:prototype}.

\subsection{Settings and Basic Characteristics}

In all of our experiments in this paper we used a Samsung Galaxy U smartphone~\cite{sgal}, which has an S5PC111 processor operating at 1GHz, 512 MB of RAM, a Gingerbread OS (Android OS 2.3), and a 5.0 Mega pixel camera with auto-focus, geo-tagging, touch focus, and face and smile detection. 
To send encrypted data from the terminal to the smartphone conveniently, we adopted the QR barcode. However, 
the camera on the phone is not error-free when it captures a QR barcode, we need to correct errors at the time of
decoding. Fortunately, QR barcode has internal error correction capability with Reed Solomon code. 
The rate of successful decoding of QR barcodes is dependent upon multiple factors such as 
a QR barcode size, the resolution of a camera and the smartphone screen. Thus, the amount of data needed to execute our protocol is crucial to its feasibility. 

If we use alphanumeric passwords only, the total number of possible keys in the onscreen keyboard is 36.
So, $\log_2{36!} = 138$ bit entropy is guaranteed which is high enough for security. 
For the encoding of this permutation, we need $36 \times \log_2{36} = 187$ bits. But if we use alphanumeric encoding of 
QR code, we need to store 36 alphanumeric characters. Considering that one QR barcode can hold up to 2953 binary characters (bytes), it is feasible to encode the keyboard arrangement in a QR barcode. More specifically, QR barcode version 2 with M level 
error correcting capability can hold 36 alphanumerics. So, even a relatively small capacity QR barcode with 
moderate error correcting code is applicable. Accordingly, we can run our authentication protocols
with a smartphone with a low-resolution camera and screen. The rate of successful scanning and 
decoding of QR challenges depends partly on the QR barcode size, and translates to higher screen time when failure happens.

\begin{figure}[htbp]
\begin{center}
\includegraphics[width=0.45 \textwidth]{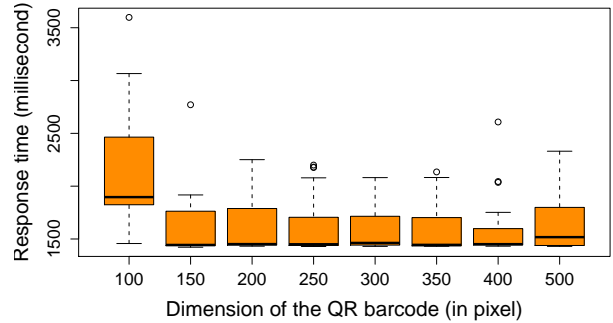}
\caption{Boxplot of the different measurements of the response time for different QR barcode size (in pixel, 20 measurements for each case). Note that the boxplot represents the 5\%, 95\% maximum, mean, first and second quartiles of the time measurements. Notice the positive association between the time and the code size.}\label{fig:resp}
\end{center}
\end{figure}

To understand the impact the QR barcode size on the time required for decoding the message using our application, we run the following experiment. Using the same QR barcode for the same encoded data, we generate QR barcodes with different sizes (in pixel). We use 8 sizes, as shown in Figure~\ref{fig:resp}. In all measurements, we demonstrate that the response time of the QR scanner is well less than 3 {\em sec}. Furthermore, for the majority of the cases, we obtain a response mean time around 1.5 seconds, except when the QR barcode size is 100 pixel, which translates into an average of 1.75 seconds. This measurement demonstrates the feasibility of our protocol's utilization of the QR barcodes for input and visualization.

\subsection{User and Usability Study}
To understand the usability aspects of our protocols, we perform a user study in several settings. We asked $20$ different users to use our system. We limit ourselves to a fully functioning implementation of the first protocol. We note that results reported in this study can be used as an upper bound on the performance of our protocols, since other protocols require less user intervention than the first protocol. We compare this case of our functioning protocol to two control groups (ground truth) of usage. The first case study considers the inputting method in typical devices using a qwerty keypad on the screen to correspond to the response time of typical password inputting time (without using our protocols or their security guarantees). The second case is the control group where we use a randomized keyboard on the terminal, which is already used as a security mechanism (see section~\ref{sec:introduction}). Accordingly, in the second case users do not need to map the randomized keyboard to a layout on the terminal---they rather directly use the one on the terminal. In both cases, the user enters his password using mouse clicks.

In all experiments, we use two password lengths $4$ and $8$ characters. Each user to use the password of his own choice. We repeat the experiment with each user for $10$ times. In the following, we summarize the main results.

\begin{figure}[htb]
\centering
\subfigure{\label{fig:4}\includegraphics[height=3.2cm]{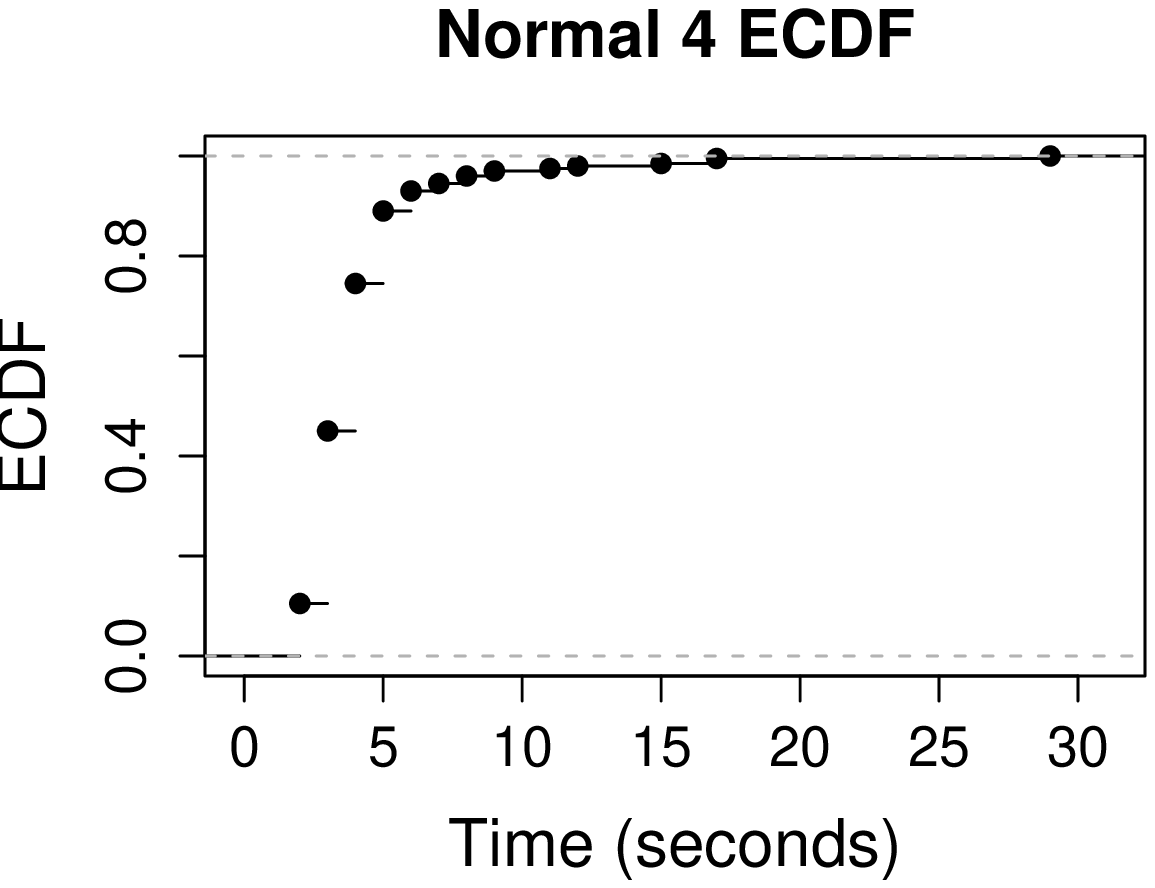}}
\subfigure{\label{fig:8}\includegraphics[height=3.2cm]{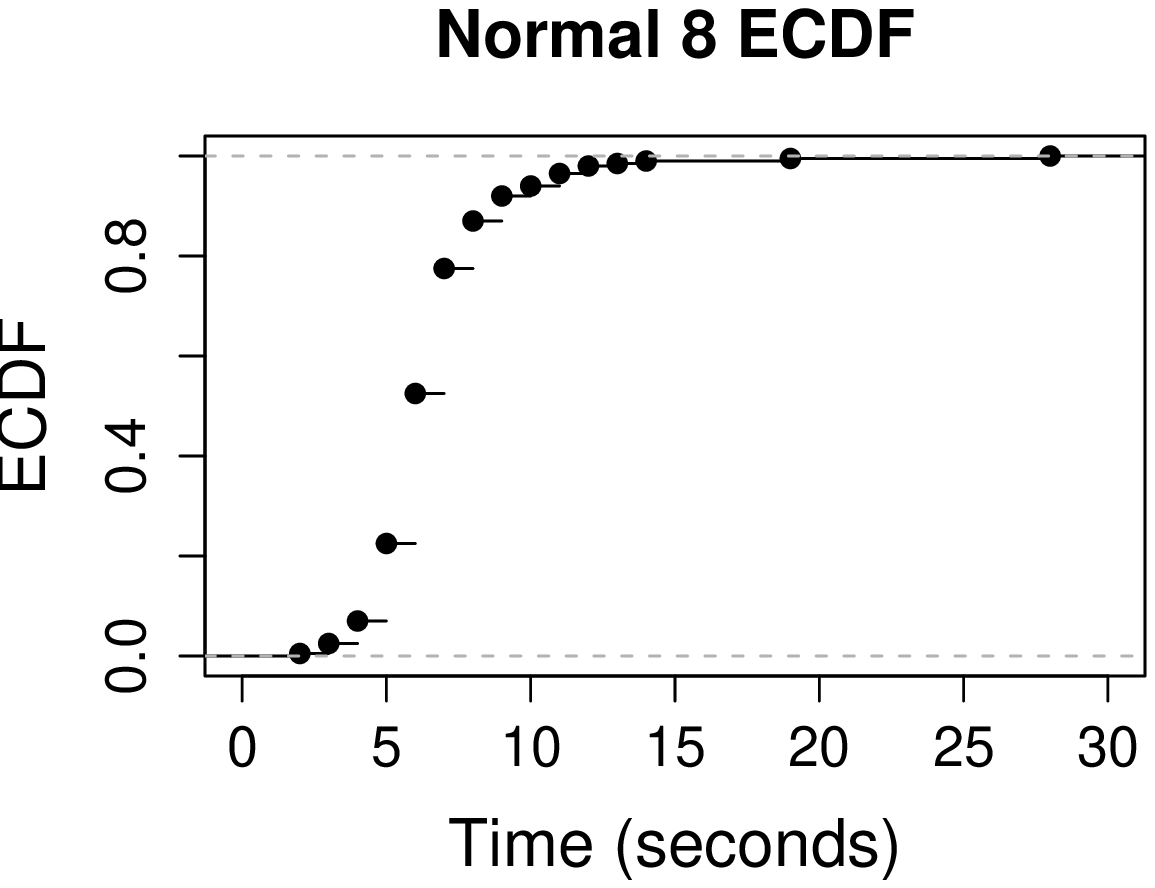}}
\caption{Empirical CDF of the time it takes for inputting passwords of two different lengths. The total number of trials is 200 obtained from 20 users. The keyboard used for input is a qwerty keypad, which is the typical Android phone keypad.}
\label{fig:normal}
\end{figure}

\parag{Case study 1---Control group (a)} 
In the first case study (referred to as {\em normal} experiment), we asked the 20 users of our system to input passwords of their own choice with the different lengths (4 and 8 characters) on a qwerty keypad shown on the terminal . Input is done by mouse clicks. We measure the response time of each user in both cases. We repeat the experiment for $10$ times for each user. We do the measurements for all cases, including  when passwords are typed incorrectly. We find that the average success rate is $97\%$ with 4 characters passwords and $91\%$ for 8 characters passwords. An empirical CDF of the time measurements (total of 200) is shown in Figure~\ref{fig:normal} (the left figure is for 4 characters and the right figure is for the 8 characters case). We find that the mean, min, max and median (in seconds) are $4.21$, $2$, $29$, and $4$ when using 4 characters passwords and $6.81$,  $2$, $28$, and $6$ when using 8 characters.

\parag{Case study 2---Control group (b)} In the second experiment (referred to as {\em randomized} experiment), we asked the same set of users to use the same set of passwords that they used in the first experiment and experimented with randomized keyboards that are rendered on the terminal. We use the same number of users and trials per user as in the previous case study. At each time, a random keyboard layout is generated, and the user is allowed to input his password to the system using mouse clicks. We note that this experiment resembles similar techniques used for defending keyloggers (see section~\ref{sec:introduction}). Same as in the previous experiment, we measure the time it takes each user to input the password using this method. We find that the average success rate is $100\%$ with 4 characters passwords and $94\%$ with $8$ characters passwords.  We plot the empirical CDF in Figure~\ref{fig:randomized} (the right side is for 4 characters and the left side is for 8 characters case) of the time measurements. We find that the mean, min, max, and median (in seconds) are $9.37$, $3$, $42$, and $8$ for length 4 and $14.73$,  $6$, $46$, and $14$ for length 8 passwords, respectively. We note that, at average, the time it takes to input passwords using this method is twice as much as when using the qwerty keypad.

\begin{figure}[htb]
\centering
\subfigure{\label{fig:4}\includegraphics[height=3.2cm]{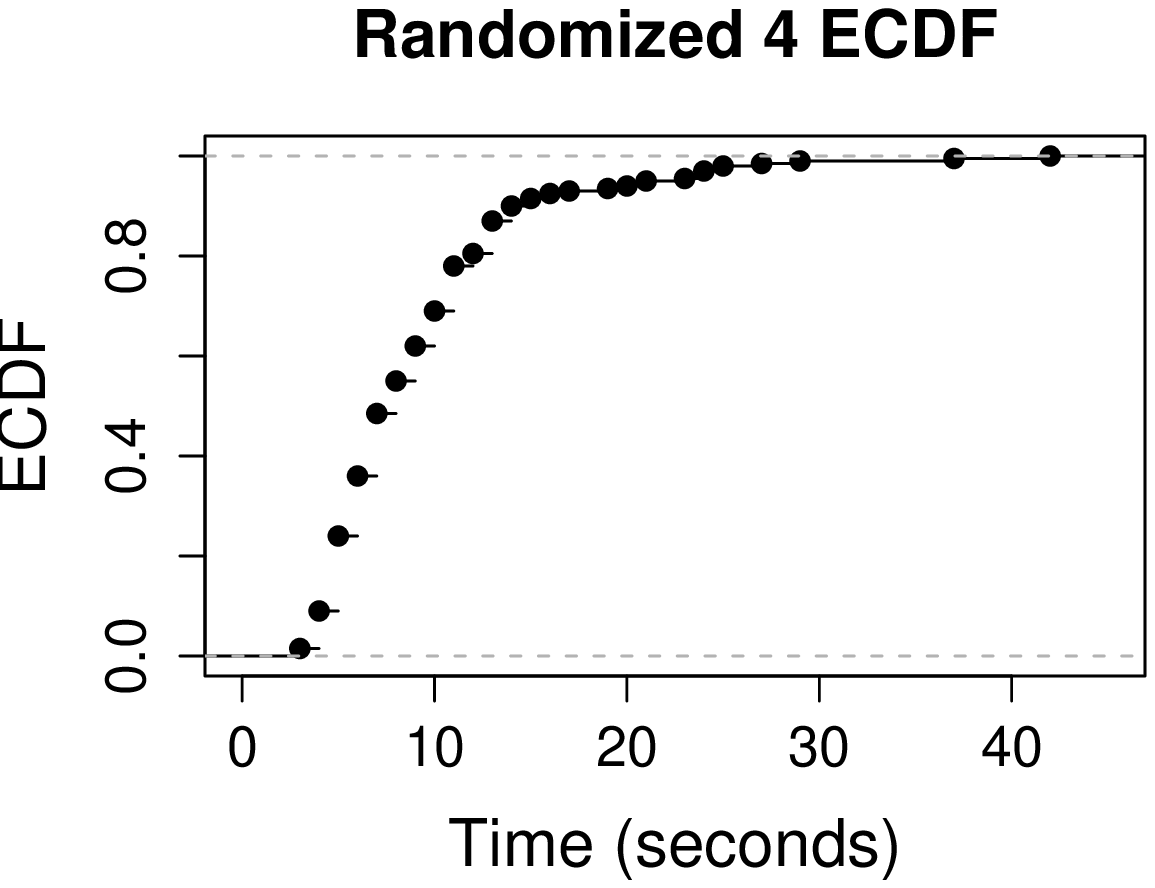}}
\subfigure{\label{fig:8}\includegraphics[height=3.2cm]{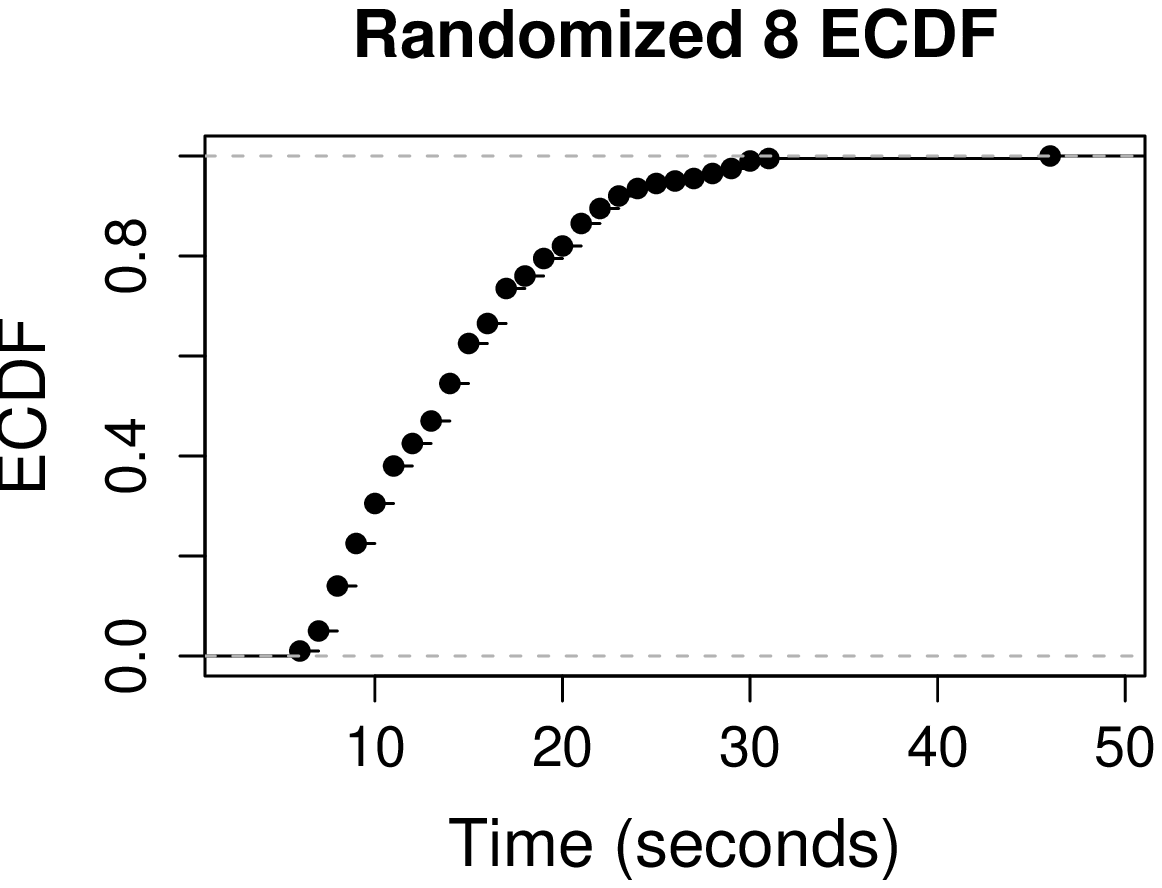}}
\caption{Empirical CDF of the time it takes for inputting passwords of two different lengths. The total number of trials are $200$ obtained from $20$ users.  The keyboard used for input is the a randomized Android phone keyboard (same keyboard is rendered on the smartphone and terminal).}
\label{fig:randomized}
\end{figure}

\parag{Case study 3---Our protocol} In this experiment, we use a fully functioning implementation of our first protocol to demonstrate its usability aspects. We use the same settings as in the prior two experiments.  In particular, the same sets of passwords with the previous control groups are used in this experiment. 
In our protocol, at each time a simulated server introduces a randomized keyboard to the user encoded into a QR barcode and encrypted as explained in section~\ref{sec:arauth}. The user is asked to input his password on the terminal's keyboard using mouse clicks with the help of the keyboard on the smartphone. Each user repeats that for 10 times, and at each time the server generates a new keyboard. We further account for all processing and computations in our protocols by timing the overall login process. We measure the overall login time since the server introduces the randomized keyboard till the user logs in (or being rejected for password mismatch). Same as before, we measure the response time even when the login fails for password mismatch. For that, we notice that the average achieved success rate with our protocol for login is $98.5\%$ for $4$ characters passwords and $96\%$ for $8$ characters passwords case. We notice that our system achieves a comparable success rate to that of both control groups supporting its usability.

\begin{figure}[htb]
\centering
\subfigure[{$4$ characters password}]{\label{fig:4}\includegraphics[height=2.5cm]{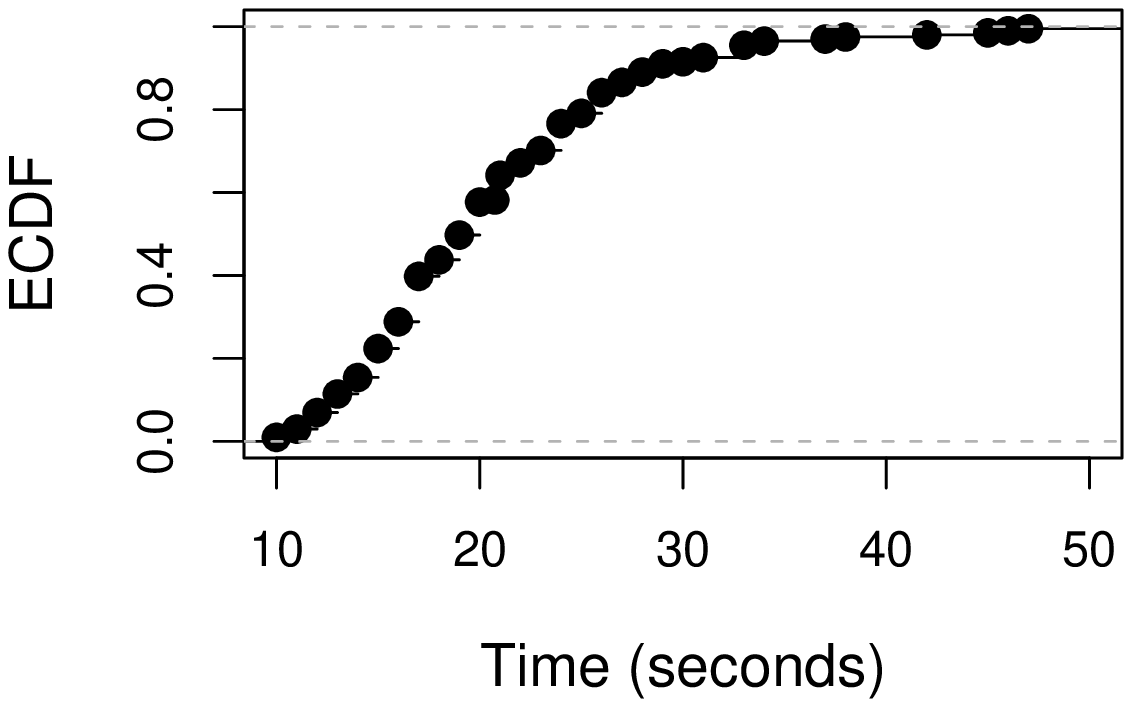}}
\subfigure[{$8$ characters password}]{\label{fig:8}\includegraphics[height=2.5cm]{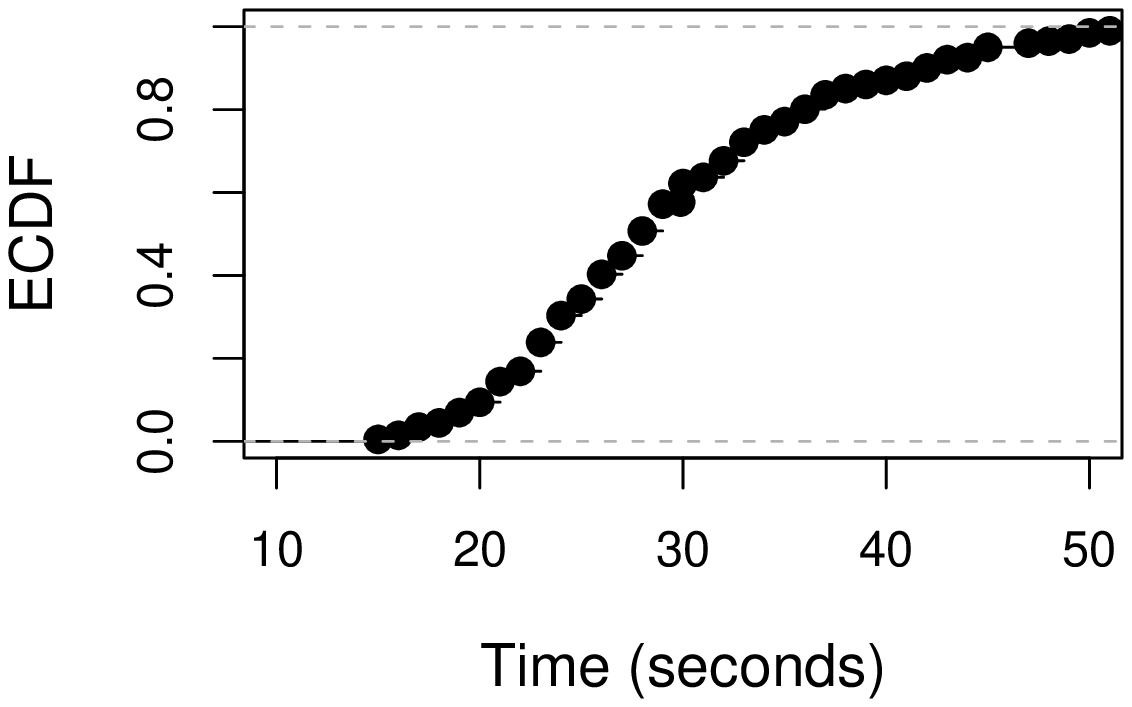}}
\caption{Empirical CDF of the time it takes for inputting passwords of two different lengths in our first protocol. The total number of trials are $200$ obtained from $20$ users.}\label{fig:userstudy}
\end{figure}

Same as before, we plot an empirical CDF of the time measurements in Figure~\ref{fig:userstudy} (detailed statistics and subjects are inTable~\ref{tab:userstudy}). The time measured in both figures includes the total time it takes for cryptographic operations, encoding, decoding, communication (negligible), and user response. We notice that our system is practical compared to the other case studies, since the mean, minimum, maximum, and median times it takes (in seconds) are $20.745$, $10$, $53$, and $19.5$ with 4 characters password and $29.81$, $15$, $52$,  and $28$ with $8$ characters (the mean $\pm 1.38$ {\em sec} for confidence interval of $95\%$).  Compared to the two other case studies, which do not provide the same security guarantees of our protocol, we find that our protocol takes at average twice as much as the randomized keyboard method and roughly four times (about fives times for passwords with length $4$) as much as the normal method (case study 1) both passwords lengths.

\begin{table}[htb]
\centering
\caption{Results of the user study with our protocol. $t(4)$ and $t(8)$ are the average time (in seconds) for each user (row) to input a password of the given length. The keyboard on the smartphone is re-randomized at each time.}\label{tab:userstudy}
\begin{tabular}{ccccc}
\toprule
$t(4)$ & success & $t(8)$ & success & gender (age) \\
\midrule
25.2		& \baa	&44.5		&\bab\ca		&male		(32)\\ \hline
27.6		& \baa	&37.6		&\baa		&male		(31)\\\hline
21.1		& \baa		&29.7		&\baa		&male 		(29)\\\hline
18.5		& \baa	&24.9		&\baa		&male		(30)\\\hline
25.3		& \baa		&39.3		&\baa		&male 		(29)\\\hline
14.8		& \baa		&23.6		&\bab\ca		&male		(27)\\\hline
16.9		& \baa		&30.6		&\baa		&male		(28)\\\hline
16.8		& \baa		&23.6		&\baa		&male		(30)\\\hline
13.1		& \baa		&34.4		&\baa		&male		(24)\\\hline
24.6		& \baa		&39.9		&\baa		&male		(29)\\\hline
28.1		& \baa		&30.5		&\baa		&male		(27)\\\hline
23.0		& \bab\ca		&31.4		&\baa		&male		(29)\\\hline
23.0		& \baa		&18.7		&\baa		&male		(25)\\\hline
20.0		&\baa		&26.5		&\baa		&female		(23)\\\hline
24.1		&\baa		&25.2		&\baa		&male		(29)\\\hline
24.0		&\baa		&26.7		&\bac\ca\ca		&male		(31)\\\hline
18.3		&\bab\ca		&25.7		&\baa		&male		(29)\\\hline
16.9		&\bab\ca		&35.0		&\bac\ca\ca		&male		(27)\\\hline
16.7		&\baa		&24.1		&\bac\ca\ca		&male		(30)\\\hline
16.9		&\baa		&25.9		&\baa		&male 		(30)\\\hline
$20.7$& $98.5\%$ & $29.9$ & $96\%$ & 1:19 ($28.7$)\\
\bottomrule
\end{tabular}
\end{table}

\section{Related Work}\label{sec:related}
There has been a large body of work on the problem of user authentication in general~\cite{lamport1981password,otway1987efficient,krawczyk1997hmac,naor1997visual,steiner1988kerberos}, and in the context of e-banking as well. Of special interest are authentication protocols that use graphical passwords like those reported in~\cite{suo2005graphical,davis2004user,thorpe2007human,Hayashi:2008:UYI} and attacks on them reported in~\cite{chiasson2008graphical,HKW06,Gao:2008:YYG,Karlof:2007:DPA,Hayashi:2008:UYI,novoa2006virtual}. To the best of our knowledge, our protocols are the first of their own types to use visualization for improving security and usability of authentication protocols as per the way reported in this paper. 

A closely related work is ``Seeing-is-Believing'' (SiB)~\cite{DBLP:conf/sp/McCunePR05} (extended in~\cite{DBLP:journals/ijsn/McCunePR09}), which uses visual channels of 2D barcodes to resist man-in-the-middle attack in device pairing. Though we utilize similar tools by using the 2D barcodes for information representation, and the visual channel for communicating this information, our protocols are further more generic than those proposed in~\cite{DBLP:conf/sp/McCunePR05}. Our protocols are tailored to the problem settings in hand, e-banking, with a different trust and attack model than that used in~\cite{DBLP:conf/sp/McCunePR05}---which results into different guarantees as explained earlier in this paper. To prevent against phishing, Parno et al. suggested the use of trusted devices to perform mutual authentication and eliminate reliance on perfect user behavior~\cite{DBLP:conf/fc/ParnoKP06}.

Slightly touched upon in this paper are keyloggers as potential attacks for credentials stealing, which are reported in~\cite{herley2006login,Holz:2009:LMU,Slowinska:2009:PTE}, and other malware which are reported in~\cite{Stone-Gross:2009,Yin:2007:PCS}. In this paper we have shown that our protocols are secure even when one of the participants in the authentication process (the terminal or smartphone) is compromised. 
\section{Conclusion \& Future Directions}\label{sec:con}

In this paper, we propose and analyze the use of user-driven
visualization to improve security and user-friendliness of
authentication protocols. Although tailored for e-banking, and high
security environments, they can also be used in other
contexts. Moreover, we have shown three realizations of protocols that
not only improve the user experience but also resist challenging
attacks, such as the keylogger attack, the shoulder-surfing attack,
and the malware attacks. Our protocols utilize simple technologies
available in most out-of-the-box smartphone devices. Furthermore, we
developed Android application of a prototype of our protocol
demonstrates its feasibility and potential in real-world deployment
and operational settings for user authentication.

Our work indeed opens the door for several other directions that we would like to investigate as a future work. 
First, we plan to investigate the design of other protocols
with more stringent performance requirements using the same tools
provided in this work. In addition, we will study methods for improving
the security and user experience by means of visualization in other
contexts, but not limited to authentication. Finally, reporting on user studies
that will benefit from a wide deployment and acceptance of our
protocols would be a parallel future work to consider as well.

\bibliographystyle{abbrv} 
\bibliography{ref}

\begin{thebibliography}{10}

\bibitem{qrcode}
{---}.
\newblock {BS ISO/IEC 18004:2006.} information technology. automatic
  identification and data capture techniques.
\newblock {ISO/IEC}, 2006.

\bibitem{qrapp}
{---}.
\newblock Qr-app.
\newblock \url{http://qr-app.appspot.com/}, 2011.

\bibitem{redlaser}
{---}.
\newblock {RedLaser}.
\newblock \url{http://redlaser.com/}, July 2011.

\bibitem{sgal}
{---}.
\newblock {Samsung Galaxy U}.
\newblock \url{http://tiny.cc/gxydt}, 2011.

\bibitem{savvy}
{---}.
\newblock {ShopSavvy}.
\newblock \url{http://shopsavvy.mobi}, 2011.

\bibitem{utf8}
{---}.
\newblock {UTF8}.
\newblock \url{http://www.utf8.com/}, 2011.

\bibitem{zxing}
{---}.
\newblock {ZXing}.
\newblock \url{http://code.google.com/p/zxing/}, 2011.

\bibitem{bcscanner}
{BahnTech}.
\newblock Barcode scanner.
\newblock \url{http://www.bahntech.com/iphone/}, 2011.

\bibitem{boneh2004short}
D.~Boneh and X.~Boyen.
\newblock Short signatures without random oracles.
\newblock In {\em Proc. of EUROCRYPT}, pages 56--73, 2004.

\bibitem{chiasson2008graphical}
S.~Chiasson, P.~van Oorschot, and R.~Biddle.
\newblock Graphical password authentication using cued click points.
\newblock In {\em Proc. of ESORICS}, 2008.

\bibitem{Crites:2008:OES}
S.~Crites, F.~Hsu, and H.~Chen.
\newblock Omash: enabling secure web mashups via object abstractions.
\newblock In {\em Proc. of ACM CCS}, pages 99--108, 2008.

\bibitem{davis2004user}
D.~Davis, F.~Monrose, and M.~Reiter.
\newblock On user choice in graphical password schemes.
\newblock In {\em Proc. of USENIX Security}, 2004.

\bibitem{Dhamija:2005:BAP}
R.~Dhamija and J.~D. Tygar.
\newblock The battle against phishing: Dynamic security skins.
\newblock In {\em Proc. of ACM SOUPS}, pages 77--88, 2005.

\bibitem{doraswamy2003ipsec}
N.~Doraswamy and D.~Harkins.
\newblock {\em IPSec: the new security standard for the Internet, intranets,
  and virtual private networks}.
\newblock Prentice Hall, 2003.

\bibitem{Gao:2008:YYG}
H.~Gao, X.~Guo, X.~Chen, L.~Wang, and X.~Liu.
\newblock Yagp: Yet another graphical password strategy.
\newblock In {\em Proc. of ACM ACSAC}, pages 121--129, 2008.

\bibitem{GoldwasserMR88}
S.~Goldwasser, S.~Micali, and R.~L. Rivest.
\newblock A digital signature scheme secure against adaptive chosen-message
  attacks.
\newblock {\em SIAM Journal}, 1988.

\bibitem{android}
Google.
\newblock Android.
\newblock \url{http://www.android.com/}, 2011.

\bibitem{Hayashi:2008:UYI}
E.~Hayashi, R.~Dhamija, N.~Christin, and A.~Perrig.
\newblock Use your illusion: secure authentication usable anywhere.
\newblock In {\em Proc. of ACM SOUPS}, 2008.

\bibitem{herley2006login}
C.~Herley and D.~Florencio.
\newblock How to login from an internet caf{\'e} without worrying about
  keyloggers.
\newblock In {\em Proc. of ACM SOUPS}, 2006.

\bibitem{HKW06}
A.~Hiltgen, T.~Kramp, and T.~Weigold.
\newblock Secure internet banking authentication.
\newblock {\em IEEE Security and Privacy}, 4:21--29, March 2006.

\bibitem{Holz:2009:LMU}
T.~Holz, M.~Engelberth, and F.~Freiling.
\newblock Learning more about the underground economy: a case-study of
  keyloggers and dropzones.
\newblock In {\em Proc. of ESORICS}, pages 1--18, 2009.

\bibitem{hopper2001secure}
N.~Hopper and M.~Blum.
\newblock Secure human identification protocols.
\newblock In {\em Proc. of ASIACRYPT}, 2001.

\bibitem{aes}
R.~Housley.
\newblock {RFC3686}: Using {A}dvanced {E}ncryption {S}tandard ({AES}) counter
  mode with ipsec encapsulating security payload ({ESP}).
\newblock \url{http://www.ietf.org/rfc/rfc3686.txt}, 2004.

\bibitem{base64}
S.~Josefsson.
\newblock {RFC 4648:} the base16, base32, and base64 data encodings.
\newblock \url{http://tools.ietf.org/html/rfc4648}, 2006.

\bibitem{Karlof:2007:DPA}
C.~Karlof, U.~Shankar, J.~D. Tygar, and D.~Wagner.
\newblock Dynamic pharming attacks and locked same-origin policies.
\newblock In {\em Proc. of ACM CCS}, pages 58--71, 2007.

\bibitem{krawczyk1997hmac}
H.~Krawczyk, M.~Bellare, and R.~Canetti.
\newblock Hmac: Keyed-hashing for message authentication.
\newblock {RFC}, 1997.

\bibitem{Kumar:2007:RSU}
M.~Kumar, T.~Garfinkel, D.~Boneh, and T.~Winograd.
\newblock Reducing shoulder-surfing by using gaze-based password entry.
\newblock In {\em Proc. of ACM SOUPS}, pages 13--19, 2007.

\bibitem{kumar2007reducing}
M.~Kumar, T.~Garfinkel, D.~Boneh, and T.~Winograd.
\newblock Reducing shoulder-surfing by using gaze-based password entry.
\newblock In {\em Proc. of ACM SOUPS}, pages 13--19, 2007.

\bibitem{lamport1981password}
L.~Lamport.
\newblock Password authentication with insecure communication.
\newblock {\em Communications of the ACM}, 24(11):770--772, 1981.

\bibitem{Lim07}
J.~Lim.
\newblock Defeat spyware with anti-screen capture technology using visual
  persistence.
\newblock In {\em Proc. of ACM SOUPS}, pages 147--148, 2007.

\bibitem{DBLP:conf/sp/McCunePR05}
J.~M. McCune, A.~Perrig, and M.~K. Reiter.
\newblock Seeing-is-believing: Using camera phones for human-verifiable
  authentication.
\newblock In {\em Proc. of IEEE Symposium on Security and Privacy}, pages
  110--124, 2005.

\bibitem{DBLP:journals/ijsn/McCunePR09}
J.~M. McCune, A.~Perrig, and M.~K. Reiter.
\newblock Seeing-is-believing: using camera phones for human-verifiable
  authentication.
\newblock {\em International Journal of Security and Networks}, 4(1/2):43--56,
  2009.

\bibitem{naor1997visual}
M.~Naor and B.~Pinkas.
\newblock Visual authentication and identification.
\newblock In {\em Proc. of CRYPTO}, 1997.

\bibitem{novoa2006virtual}
M.~Novoa, V.~Ali, and M.~Altendorf.
\newblock Virtual user authentication system and method.
\newblock US Patent App. 20,080/028,441, 2006.

\bibitem{otway1987efficient}
D.~Otway and O.~Rees.
\newblock Efficient and timely mutual authentication.
\newblock {\em ACM SIGOPS Operating Systems Review}, 21(1):8--10, 1987.

\bibitem{DBLP:conf/fc/ParnoKP06}
B.~Parno, C.~Kuo, and A.~Perrig.
\newblock Phoolproof phishing prevention.
\newblock In {\em Proc. of Financial Cryptography}, pages 1--19, 2006.

\bibitem{pemmaraju2007methods}
R.~Pemmaraju.
\newblock Methods and apparatus for securing keystrokes from being intercepted
  between the keyboard and a browser.
\newblock {Patent 182,714}.

\bibitem{rescorla2001ssl}
E.~Rescorla.
\newblock {\em SSL and TLS: designing and building secure systems}.
\newblock Addison-Wesley, 2001.

\bibitem{Slowinska:2009:PTE}
A.~Slowinska and H.~Bos.
\newblock Pointless tainting?: evaluating the practicality of pointer tainting.
\newblock In {\em Proc. of ACM EuroSys}, pages 61--74, 2009.

\bibitem{steiner1988kerberos}
J.~Steiner, C.~Neuman, and J.~Schiller.
\newblock Kerberos: An authentication service for open networks.
\newblock In {\em Proc. of USENIX Annual Tech Conference}, pages 191--201,
  1988.

\bibitem{Stone-Gross:2009}
B.~Stone-Gross, M.~Cova, L.~Cavallaro, B.~Gilbert, M.~Szydlowski, R.~Kemmerer,
  C.~Kruegel, and G.~Vigna.
\newblock Your botnet is my botnet: analysis of a botnet takeover.
\newblock In {\em Proc. of ACM CCS}, pages 635--647, 2009.

\bibitem{suo2005graphical}
X.~Suo, Y.~Zhu, and G.~Owen.
\newblock Graphical passwords: A survey.
\newblock {\em IEEE Computer}, 2005.

\bibitem{thorpe2007human}
J.~Thorpe and P.~van Oorschot.
\newblock Human-seeded attacks and exploiting hot-spots in graphical passwords.
\newblock In {\em Proc. of USENIX Security}, 2007.

\bibitem{vizcaino1994method}
G.~Vizcaino.
\newblock Method and apparatus for securing credit card transactions.
\newblock {US Patent 5,317,636}, 1994.

\bibitem{Yin:2007:PCS}
H.~Yin, D.~Song, M.~Egele, C.~Kruegel, and E.~Kirda.
\newblock Panorama: capturing system-wide information flow for malware
  detection and analysis.
\newblock In {\em Proc. of ACM CCS}, 2007.

\end{thebibliography}

\appendix
\section{Experiments}
Here we report on the case studies and the time measurements, so of which are reported in section~\ref{sec:exp}. To characterize the time measurements, we use boxplots, which capture the the 5\%, 95\% maximum, mean, first and second quartiles, and the outliers of measurements. Outliers are measurements that are $1.4$ times the 95\% maximum or $1.4$ times the $5\%$ minimum.

\begin{figure}[htb]
\centering
\subfigure[4 Characters]{\label{fig:respn4}\includegraphics[width=0.4\textwidth]{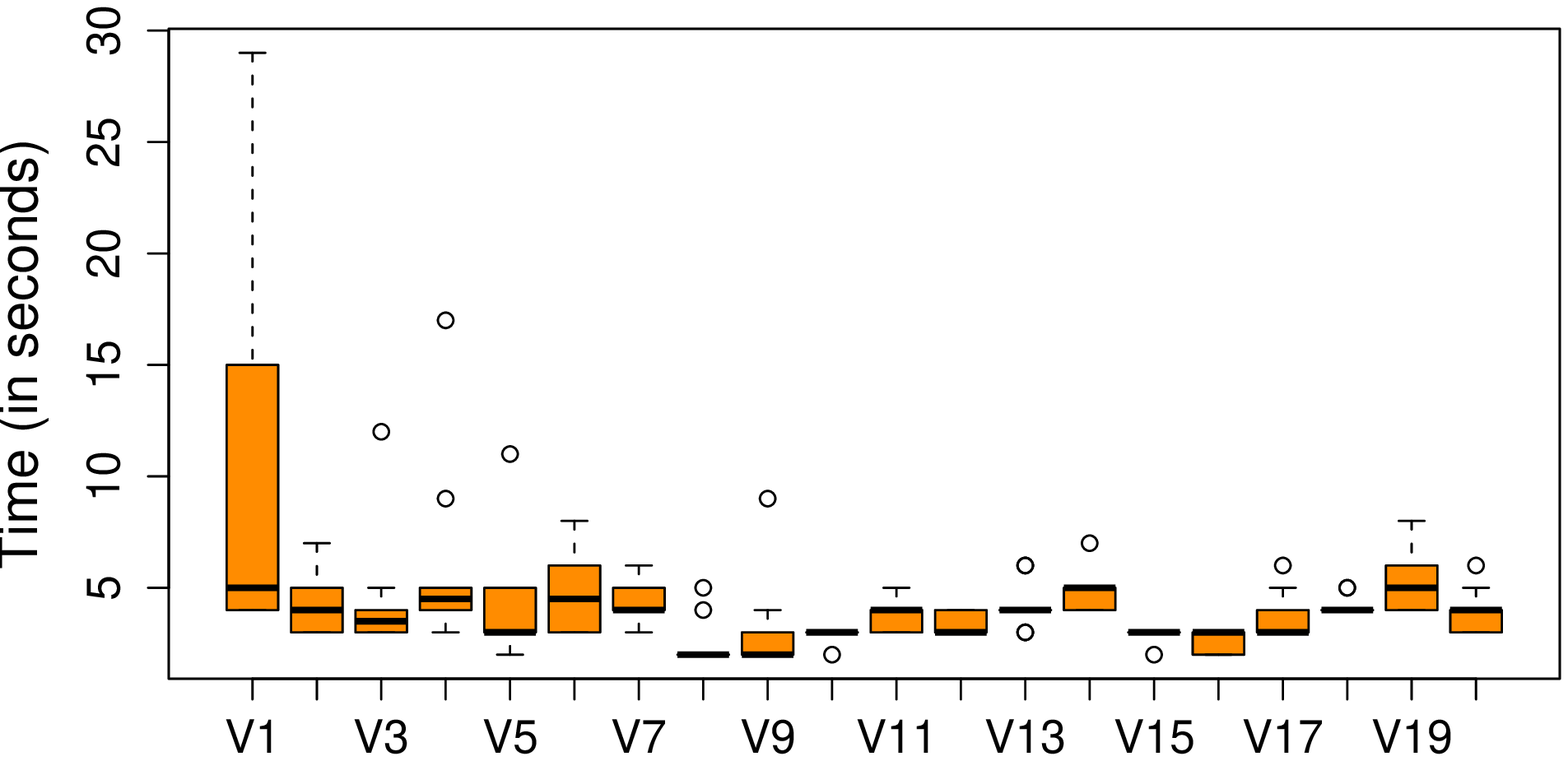}}\\ \vspace{-3mm}
\subfigure[8 Characters]{\label{fig:respn8}\includegraphics[width=0.4\textwidth]{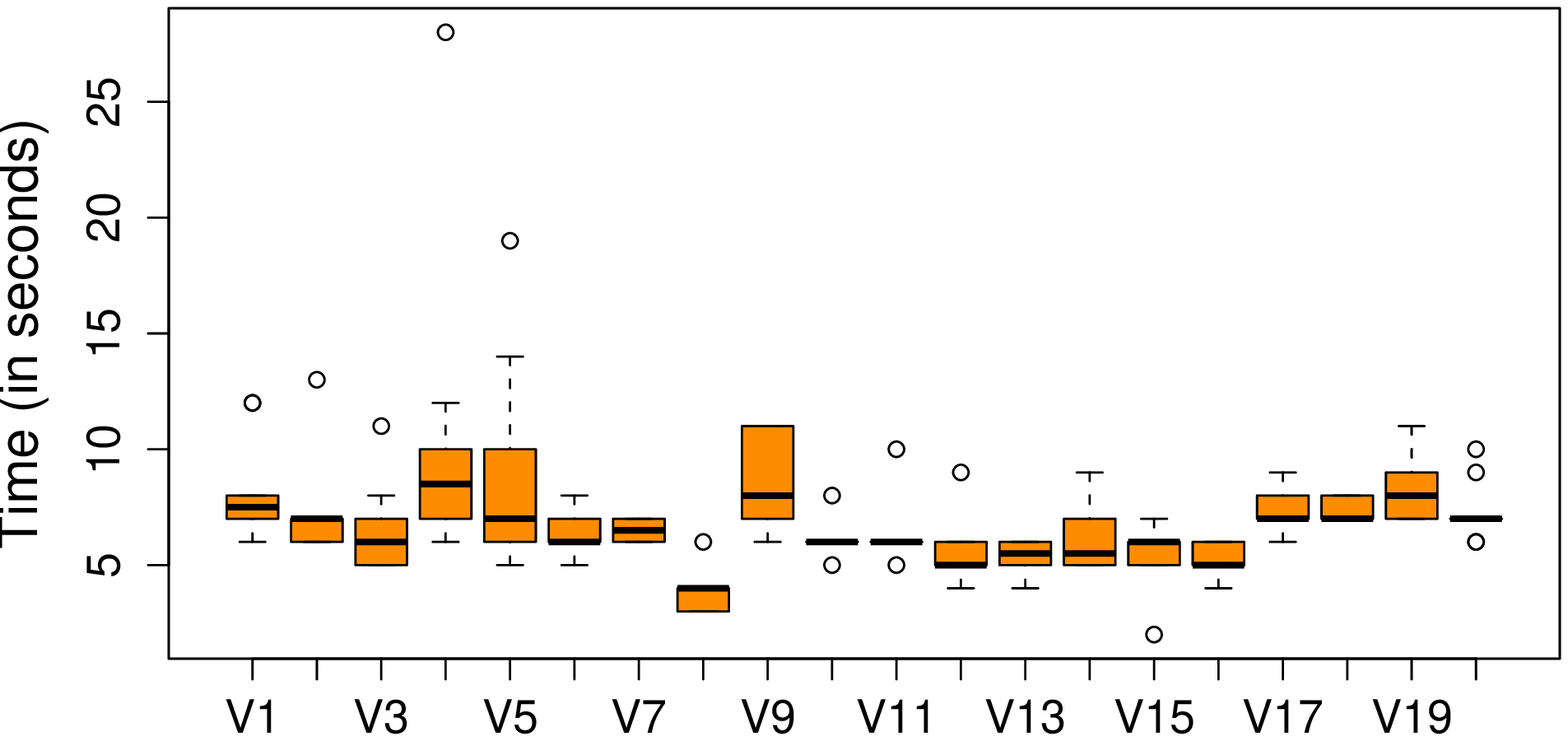}}\vspace{-3mm}
\caption{Boxplot of the different measurements of the users response time when inputting different passwords with length $4$ and $8$ characters with the case study 1. The boxplot captures the maximum (top $95\%$), the minimum (lowest $5\%$), the median, the first and third quartiles (medians of the first and second halves of the population). Note the circles are outliers (more than $1.4$ times the maximum)}
\label{fig:randomized}
\end{figure}

\parag{Case study 1} The boxplot of the raw time measures (per subject) is shown in Figure~\ref{fig:respn4} for 4 characters case and in Figure~\ref{fig:respn8} for 8 characters case. More measurements, the passwords, and error rates are shown in Table~\ref{tab:normal}.

\begin{table}[htb]
\centering
\caption{Results of the user study with qwerty keypad. The time is the average time per 10 trials (in seconds) for each user (row) to input the password. The keyboard on the smartphone is re-randomized at each time. The success is computed out of 10 trials per each user.}\label{tab:normal}
\begin{tabular}{cccccc}
\toprule
pwd (4)	&	time	&	success	&	pwd (8)	&	time	&	success	\\	\midrule
t4pw	&	9.4	&	10	&	  aplegw2j	&	8.2	&	9	\\	\hline
data	&	4.2	&	9	&	  keywords	&	7.2	&	9	\\	\hline
code	&	4.4	&	9	&	  asdfnews	&	6.4	&	10	\\	\hline
 usa8	&	6	&	10	&	  microarm	&	10.4	&	8	\\	\hline
  head	&	4.1	&	10	&	  networks	&	8.8	&	4	\\	\hline
1213	&	4.6	&	10	&	  jjae1213	&	6.4	&	10	\\	\hline
1596	&	4.2	&	9	&	  mbc2356z	&	6.5	&	10	\\	\hline
4455	&	2.5	&	10	&	 dd4455ee	&	3.8	&	10	\\	\hline
2222	&	3	&	10	&	78061622	&	8.5	&	10	\\	\hline
1317	&	2.9	&	9	&	   ucsl1317	&	6.1	&	8	\\	\hline
   bhnj	&	3.7	&	10	&	   1116baek	&	6.3	&	10	\\	\hline
1317	&	3.4	&	9	&	12341317	&	5.4	&	7	\\	\hline
712	&	4.2	&	10	&	88061100	&	5.4	&	10	\\	\hline
   aple	&	4.8	&	10	&	   hjh37267	&	6	&	9	\\	\hline
   save	&	2.9	&	10	&	20110714	&	5.5	&	9	\\	\hline
   cola	&	2.7	&	10	&	   cocacola	&	5.2	&	10	\\	\hline
   asdf	&	3.6	&	9	&	19830325	&	7.2	&	10	\\	\hline
503	&	4.2	&	9	&	   babosmjk	&	7.3	&	9	\\	\hline
   u137	&	5.4	&	9	&	   coffee20	&	8.3	&	10	\\	\hline
   uku0	&	4	&	10	&	   terminal	&	7.3	&	10	\\	\hline
\bottomrule
\end{tabular}
\end{table}

\begin{figure}[htb]
\centering
\subfigure[4 Characters]{\label{fig:respr4}\includegraphics[width=0.4\textwidth]{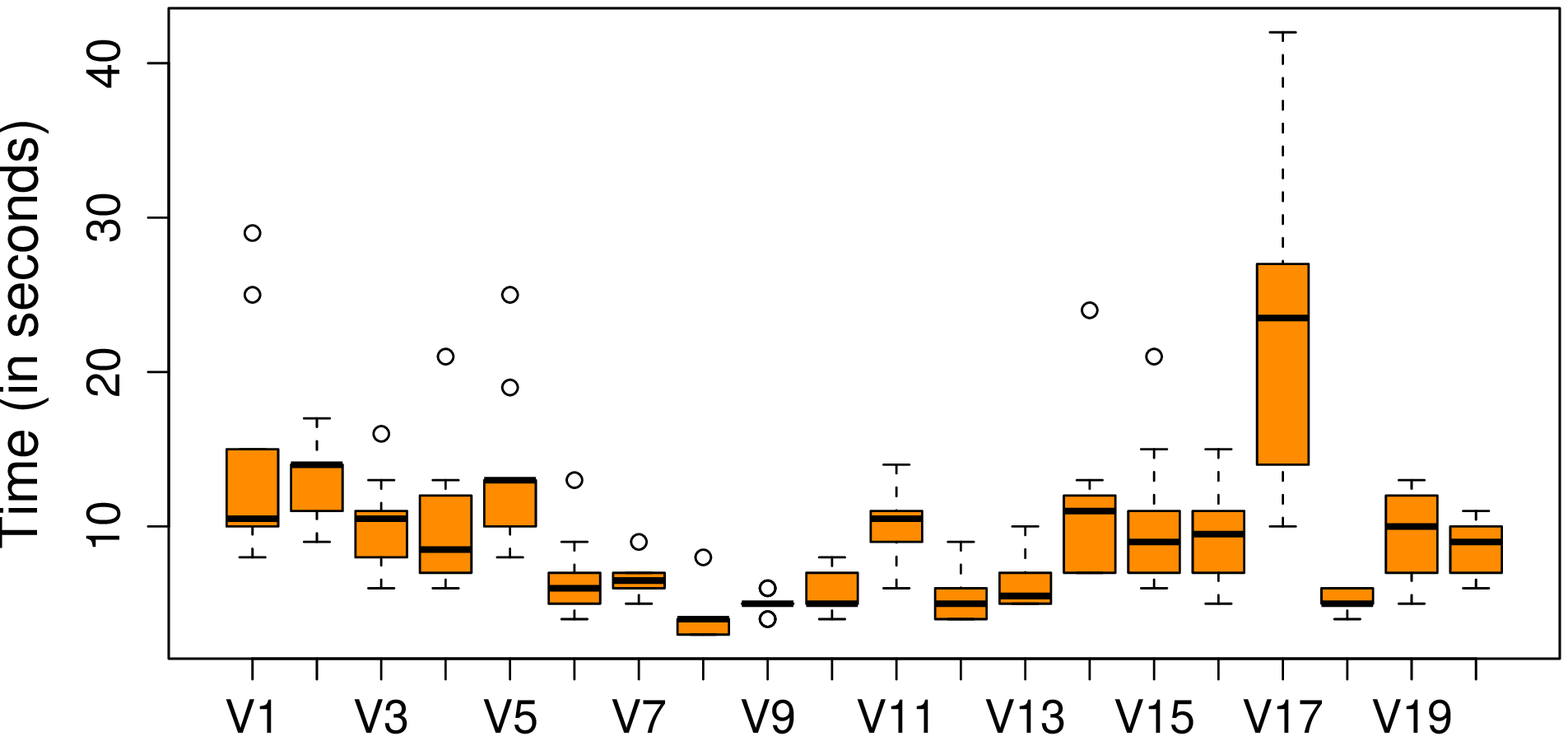}}\\ \vspace{-3mm}
\subfigure[8 Characters]{\label{fig:respr8}\includegraphics[width=0.4\textwidth]{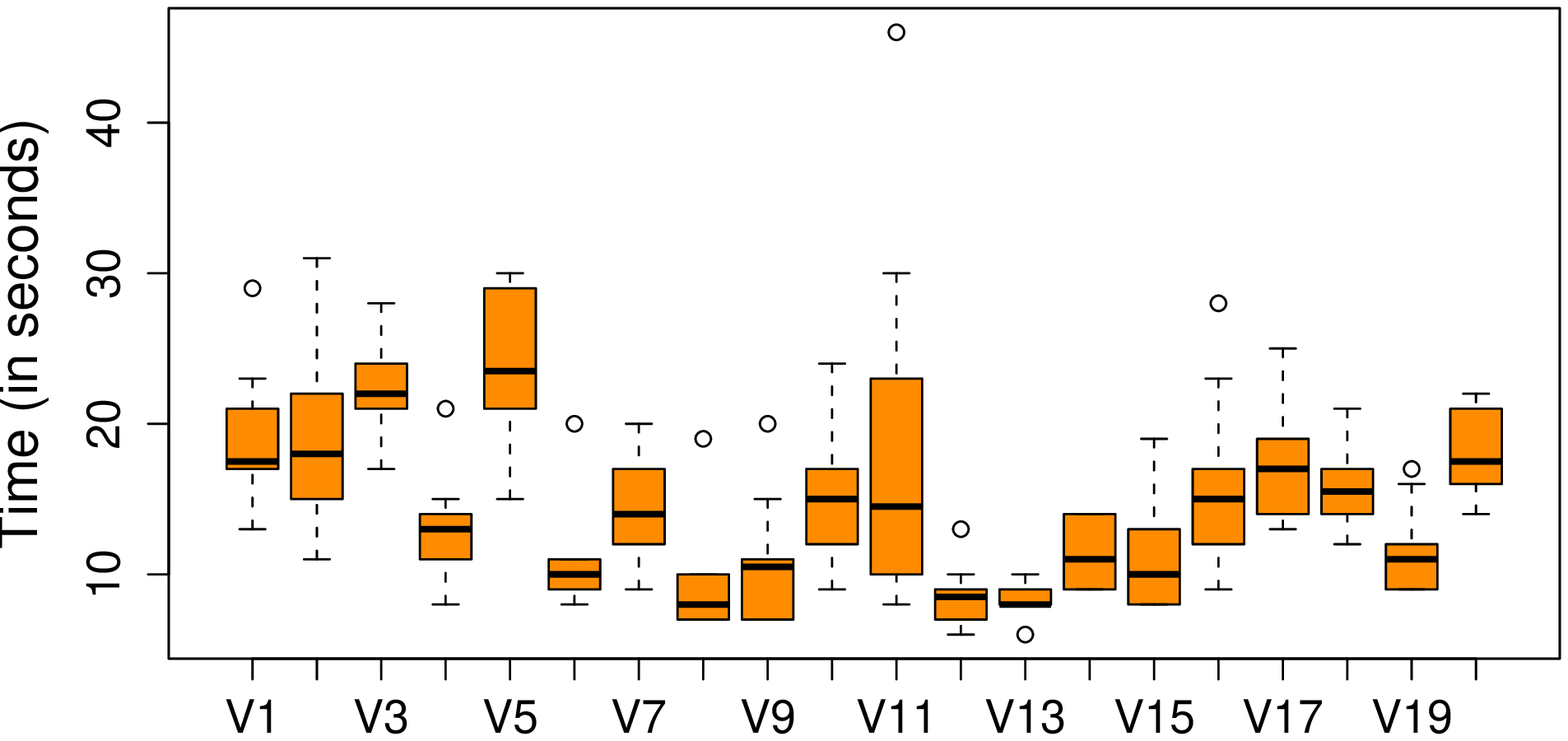}}\vspace{-3mm}
\caption{Boxplot of the different measurements of the users response time when inputting different passwords with length $4$ and $8$ characters with the case study 2.}
\label{fig:randomized}
\end{figure}

\parag{Case study 2} The boxplot of the raw time measures (per subject) is shown in Figure~\ref{fig:respr4} for 4 characters case and in Figure~\ref{fig:respr8} for 8 characters case. More measurements, the passwords, and error rates are shown in Table~\ref{tab:randomized}.

\begin{table}[htb]
\centering
\caption{Results of the user study with the randomized keyboard. The time is the average time per 10 trials (in seconds) for each user (row) to input the password. The keyboard on the smartphone is re-randomized at each time. The success is computed out of 10 trials per each user.}\label{tab:randomized}
\begin{tabular}{cccccc}
\toprule
pwd (4)	&	time	&	success	&	pwd (8)	&	time	&	success	\\	\midrule
t4pw	&	13.9	&	10	&	  aplegw2j	&	18.8	&	9	\\	\hline
data	&	13	&	10	&	  keywords	&	18.9	&	9	\\	\hline
code	&	10.1	&	10	&	  asdfnews	&	22.3	&	10	\\	\hline
 usa8	&	10.2	&	10	&	  microarm	&	13.2	&	10	\\	\hline
  head	&	13.3	&	10	&	  networks	&	24	&	10	\\	\hline
1213	&	6.7	&	10	&	  jjae1213	&	10.8	&	9	\\	\hline
1596	&	6.6	&	10	&	  mbc2356z	&	14.4	&	7	\\	\hline
4455	&	4.1	&	10	&	 dd4455ee	&	9.4	&	10	\\	\hline
2222	&	5	&	10	&	78061622	&	10.7	&	10	\\	\hline
1317	&	5.5	&	10	&	   ucsl1317	&	15.1	&	10	\\	\hline
   bhnj	&	10.1	&	10	&	   1116baek	&	18.4	&	9	\\	\hline
1317	&	5.5	&	10	&	12341317	&	8.6	&	9	\\	\hline
712	&	6.1	&	10	&	88061100	&	8.4	&	10	\\	\hline
   aple	&	11.2	&	10	&	   hjh37267	&	11.4	&	10	\\	\hline
   save	&	10.1	&	10	&	20110714	&	11.1	&	7	\\	\hline
   cola	&	9.3	&	10	&	   cocacola	&	16.1	&	10	\\	\hline
   asdf	&	23.4	&	10	&	19830325	&	17.3	&	10	\\	\hline
503	&	5.3	&	10	&	   babosmjk	&	16	&	10	\\	\hline
   u137	&	9.4	&	10	&	   coffee20	&	11.4	&	9	\\	\hline
   uku0	&	8.6	&	10	&	   terminal	&	18.3	&	10	\\	\hline
\bottomrule
\end{tabular}
\end{table}

\begin{figure}[htb]
\centering
\subfigure[4 Characters]{\label{fig:data04}\includegraphics[width=0.4\textwidth]{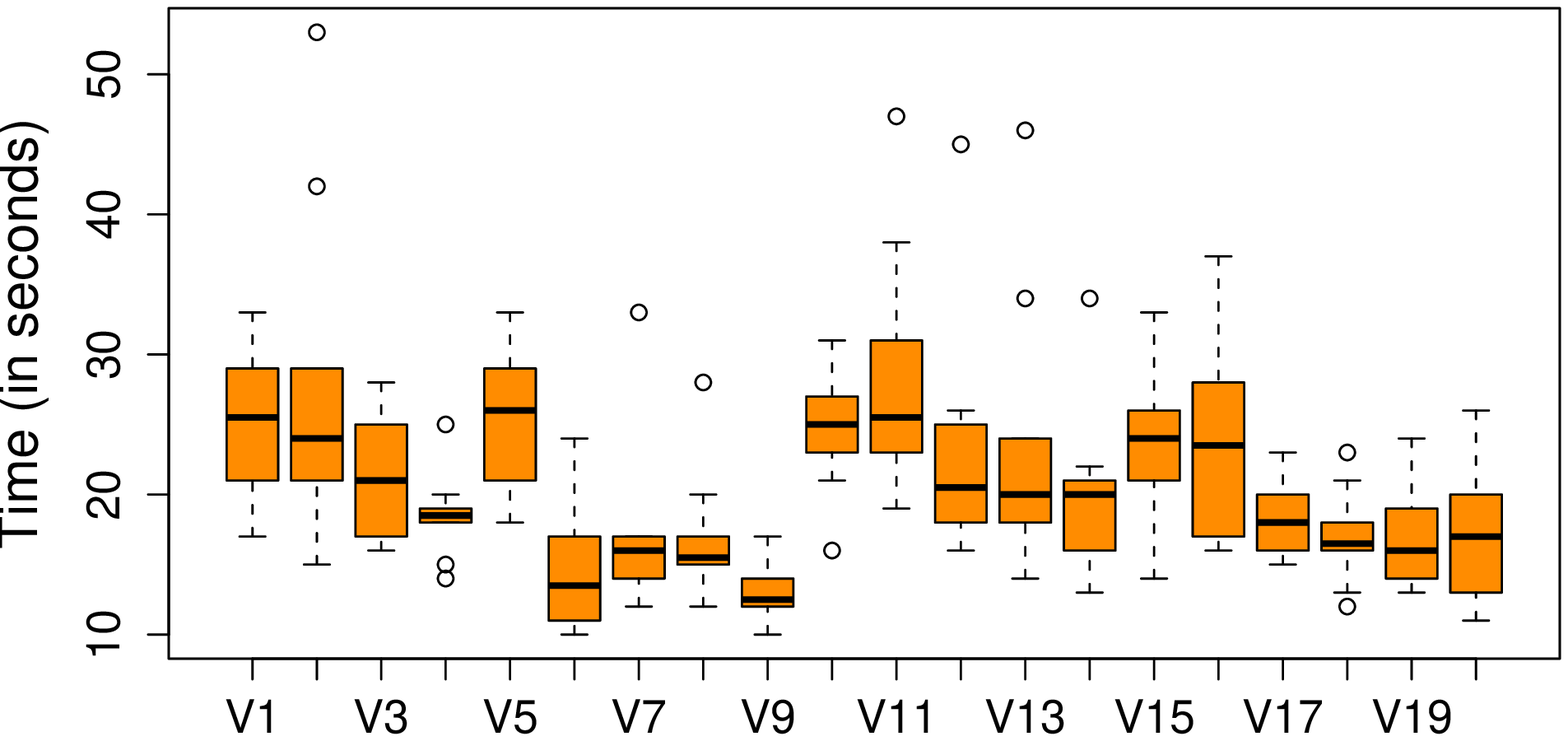}}\vspace{-3mm}
\subfigure[8 Characters]{\label{fig:data08}\includegraphics[width=0.4\textwidth]{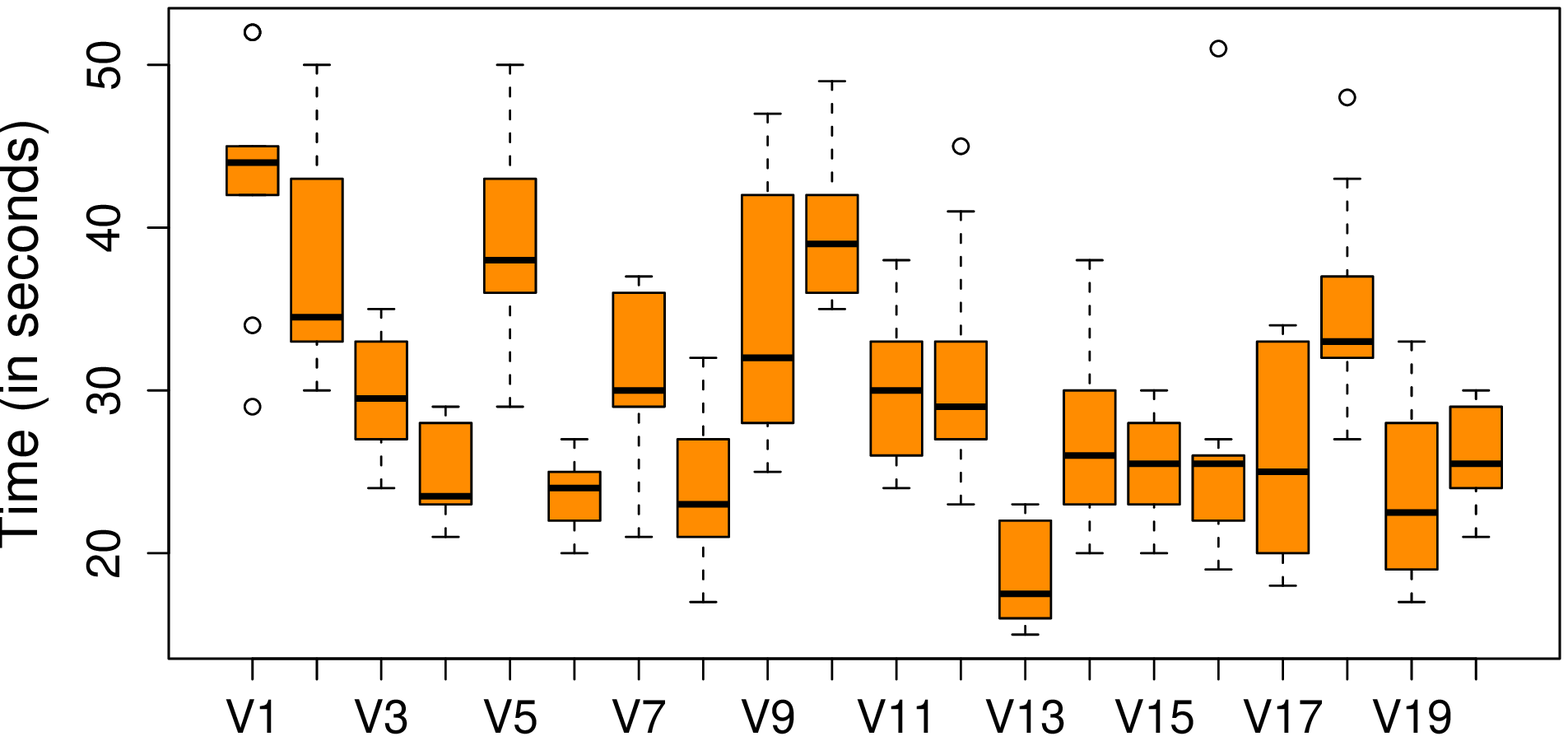}}\vspace{-3mm}
\caption{Boxplot of the different measurements of the users response time when inputting different passwords with length $4$ and $8$ characters in protocol's experiment.}
\label{fig:randomized}
\end{figure}

\parag{Case study 3} The boxplot of the raw time measures (per subject) is shown in Figure~\ref{fig:data04} for 4 characters case and in Figure~\ref{fig:data08} for 8 characters case. More measurements, the passwords, and the raw error rates are shown in Table~\ref{tab:ours}.

\begin{table}[htb]
\centering
\caption{Results of the user study with our protocol. The time is the average time per 10 trials (in seconds) for each user (row) to input the password. The keyboard on the smartphone is re-randomized at each time. The success is computed out of 10 trials per each user.}\label{tab:ours}
\begin{tabular}{cccccc}
\toprule
pwd (4)	&	time	&	success	&	pwd (8)	&	time	&	success	\\	\midrule
t4pw	&	25.2	&	10	&	  aplegw2j	&	42.9	&	9	\\	\hline
data	&	27.6	&	10	&	  keywords	&	37.6	&	10	\\	\hline
code	&	21.1	&	10	&	  asdfnews	&	29.7	&	10	\\	\hline
 usa8	&	18.5	&	10	&	  microarm	&	24.9	&	10	\\	\hline
  head	&	25.3	&	10	&	  networks	&	39.3	&	10	\\	\hline
1213	&	14.8	&	10	&	  jjae1213	&	23.6	&	9	\\	\hline
1596	&	16.9	&	10	&	  mbc2356z	&	30.6	&	10	\\	\hline
4455	&	16.8	&	10	&	 dd4455ee	&	23.6	&	10	\\	\hline
2222	&	13.1	&	10	&	78061622	&	34.4	&	10	\\	\hline
1317	&	24.6	&	10	&	   ucsl1317	&	39.9	&	10	\\	\hline
   bhnj	&	28.1	&	10	&	   1116baek	&	30.5	&	10	\\	\hline
1317	&	23	&	9	&	12341317	&	31.4	&	10	\\	\hline
712	&	23	&	10	&	88061100	&	18.7	&	10	\\	\hline
   aple	&	20	&	10	&	   hjh37267	&	26.5	&	10	\\	\hline
   save	&	24.1	&	10	&	20110714	&	25.2	&	10	\\	\hline
   cola	&	24	&	10	&	   cocacola	&	26.7	&	8	\\	\hline
   asdf	&	18.3	&	9	&	19830325	&	25.7	&	10	\\	\hline
503	&	16.9	&	9	&	   babosmjk	&	35	&	8	\\	\hline
   u137	&	16.7	&	10	&	   coffee20	&	24.1	&	8	\\	\hline
   uku0	&	16.9	&	10	&	   terminal	&	25.9	&	10	\\	\hline
\bottomrule
\end{tabular}
\end{table}
\end{document}